\documentclass[fleqn,usenatbib]{mnras}

\usepackage{newtxtext,newtxmath}

\usepackage[T1]{fontenc}
\usepackage{ae,aecompl}

\usepackage{hyperref}
\usepackage{graphicx}	
\usepackage{blindtext}  
\usepackage{natbib}
\usepackage{rotating,times,graphicx,latexsym}
\usepackage{color}
\usepackage{longtable}
\usepackage{lscape}
\usepackage{lipsum} 
\usepackage{array}
\usepackage{flafter}

\usepackage{orcidlink}

\usepackage{soul} 
\usepackage{xargs} 

\graphicspath{ {images_pdf/} }

\title[YSO Light Curve Comparison]{A survey for variable young stars with small telescopes: X - Comparing stochastic YSO light curves}

\author[Benjamin W. Ryan et al.]{Benjamin W. Ryan$^{1}$\thanks{E-mail: bwr2@kent.ac.uk}, 
Holly Stokes-Geddes$^{1}$\thanks{E-mail: hs442@kent.ac.uk}
Dirk Froebrich\orcidlink{0000-0003-4734-3345}$^{1}$\thanks{E-mail: D.Froebrich@kent.ac.uk}
\\
$^{1}$Centre for Astrophysics and Planetary Science, School of Physics and Astronomy, University of Kent, Canterbury CT2 7NH, UK\\
}

\date{Accepted XXX. Received YYY; in original form ZZZ}

\pubyear{2024}

\begin{document}
\label{firstpage}
\pagerange{\pageref{firstpage}--\pageref{lastpage}}
\maketitle

\begin{abstract} 

Light curves of young stars exhibit photometric variability over hours to decades and across a wide range of amplitudes. On time scales beyond a few rotation periods, these light curves are typically stochastic. The variability arises from a combination of accretion rate changes, line-of-sight extinction variations, and evolving spotted stellar surfaces. We aim to develop a methodology to quantitatively compare the full variability statistics of these inhomogeneously sampled light curves with model calculations. To achieve this, we converted the light curves into variability fingerprints. They map the probability of variation by a given amount over a given timescale. Applying principal component analysis to these fingerprints produces a stable distribution of the first two principal components. We show that this distribution is a continuum without clusters. Adding a model-generated fingerprint to an observational sample does not significantly alter the distribution of the sample, allowing a robust comparison between the model and observed light curves to assess statistical realism. We show that photometric uncertainties, timing, and observing cadence have a minimal impact on model placement within the observational distribution. The main source of variance among highly variable light curves of young stars is the timescale of the onset of significant variability (above 0.3~mag), with 1~–~3~month timescales being the most critical. The secondary cause of variance are long-term (above 1.5~yr) dimming or rising trends.

\end{abstract}



\section{Introduction}\label{intro}

Photometric variability has long been known as a defining feature of young stars \citep{1945ApJ...102..168J}. This prevalence of variability in young stellar objects (YSOs) is caused by a variety of physical reasons such as rotation and surface spots, variations in mass accretion rates, and changes in the line of sight extinction due to orbiting disk material. Typically, light curves of YSOs are influenced by a combination of all of these effects simultaneously, creating stochastic variability in particular at timescales above a few rotation periods, which can make it difficult to easily disentangle the effects of the different physical causes. Furthermore, the variations in brightness can reach up to several magnitudes and occur on timescales from days to tens of years. Thus, numerous monitoring campaigns have been devoted to variability studies of YSOs in the past. See e.g. the Protostars and Planets reviews by \citet{2007prpl.conf..297H, 2007prpl.conf..479B, 2014prpl.conf..433B, 2014prpl.conf..387A, 2023ASPC..534..355F} for an overview of some aspects of this vast field.

To study the underlying causes of the YSOs variability using photometry, we hence must cover all the relevant timescales and observe a statistically significant number of objects. Furthermore, observations in multiple filters are paramount to potentially distinguish variations caused by different physical mechanisms. There are multiple studies available that cover some of these aspects \citep[high cadence, long-term observations, multi filter coverage, large sample, e.g.][]{2007A&A...461..183G, 2014AJ....147...82C, 2017PASP..129j4502K}, but not all simultaneously. We have hence started the Hunting Outbursting Young Stars (HOYS) project in 2014 \citep{2018MNRAS.478.5091F} with the aim of monitoring a large number of YSOs in multiple optical filters with a daily cadence over several decades. 

Some YSO light curves can (in part) show periodic modulation due to rotation \citep[see reviews by][]{2007prpl.conf..297H, 2014prpl.conf..433B} or orbiting warped inner disk material \citep[e.g.][]{1999A&A...349..619B}. While these regular variations are common at shorter timescales \citep[e.g.][]{2022AJ....163..212C}, at longer timescales the light curves are typically stochastic. Thus, the characterisation of them is non-trivial. Multiple attempts have been made in the past to characterise light curve properties with simple metrics such as variability indices \citep[e.g.][]{1996PASP..108..851S, 2017MNRAS.464..274S}, or quasiperiodicity (Q) and asymmetry (M) indices \citep[e.g.][]{2014AJ....147...82C, 2022MNRAS.514.2736L}. While these are easy to calculate and compare (to each other or with models) for non-homogeneously sampled light curves, they essentially reduce the vast information of the entire light curve to a single number / dimension. Thus, they cannot capture the full extent of the variability information available for each YSO. Furthermore, it is not clear whether they represent the best way to distinguish YSO light curves from each other. 

There are attempts to remedy this situation \citep[e.g.][]{2004A&A...419..249S, 2015ApJ...798...89F, 2017MNRAS.465.3889R}, upon which we have built our variability fingerprinting of light curves \citep{2020MNRAS.493..184E, 2022MNRAS.510.2883F}. These fingerprints convert the light curves into a two dimensional map showing the probability that a star varies by a given magnitude on a given timescale. They can hence capture the entire range of variability of each light curve. To some extent, the fingerprints are similar to the analyses of structure functions presented e.g. by \citet{2020MNRAS.491.5035S} or \citet{2022MNRAS.514.2736L}. However, they distinguish between dimming and brightening more comprehensively and will be applied to data taken over much longer timescales (10~yr) with nearly continuous coverage in our work.

These variability fingerprints can also be created from simulated light curves based on numerical models of the evolving young stars and their surrounding disks. To determine whether the variability in the simulated light curves is statistically similar to the observed light curves of YSOs, we need to develop a method to quantitatively compare samples of these fingerprints. Unsupervised machine learning techniques, such as dimension reduction and clustering with DBSCAN or k-means might provide a method to achieve this. However, tests need to be performed to verify if these methods result in individual clusters of objects or if they create a continuum distribution. Furthermore, we need to examine how the various methods perform if samples of real data are investigated together with model fingerprints. Ideally, we are seeking to develop a methodology that creates clusters from a sample of fingerprints and is stable and robust. In other words, adding individual fingerprints based on model light curves into the sample to show where amongst the observed objects the model falls should not significantly change the assignments of the fingerprints to their clusters, or their relative positions within a continuum distribution.

In this work, we describe the data calibration and verification as well as the selection of our highly variable YSO sample in Sect.~\ref{sect_data}. In Sect.~\ref{sect_fingerprints} we lay out the determination of the variability fingerprints and discuss their accuracy. Finally, Sect.~\ref{sect_clustering} discusses the methodology and results of fingerprint analysis, as well as its stability, when comparing model and observed light curves.

\section{Photometry Data and Variable Star Selection}\label{sect_data}

In this section, we will describe the photometry data and their calibration, as well as the selection of variable stars for our analysis.

\subsection{HOYS Data and calibration}\label{data_calib}

All photometry data used are provided by the HOYS citizen science project \citep{2018MNRAS.478.5091F}. It aims to provide long-term, high cadence, multi-filter, optical photometry of 25 young (age $< 10$~Myr), nearby (distance $<1$~kpc) star forming regions. The data are collected by a mixture of $\sim$130 professional, university and amateur observatories. However, data taking for most of the light curves analysed is dominated by three 40~cm class observatories: i) the University of Kent's Beacon observatory \citep{2018MNRAS.478.5091F}; ii) the AstroLAB IRIS Observatory \citep{2020MNRAS.493..184E}; iii) the Remote Observatory Atacama Desert \citep[ROAD,][]{2012JAVSO..40.1003H, 2022MNRAS.510.2883F}. Collectively, they currently account for two thirds of all images and just over half of the total exposure time.

All images for the project are bias, dark, and flat corrected, uploaded to our publicly available Web server\footnote{\tt \url{https://astro.kent.ac.uk/HOYS-CAPS/}}, and preliminary calibrated. The astrometry in the images is solved using the {\tt Astrometry.net} software \citep{2008ASPC..394...27H}. Aperture photometry is performed using the Source Extractor software \citep{1996A&AS..117..393B}. Deep images of all regions obtained under photometric conditions in $u$, $B$, $V$, $R_c$, and $I_c$ filters are used as reference for relative photometry. In this work we only investigate data taken in the filters with most data, which are referred to as $V$, $R$, and $I$, hereafter. This follows the steps outlined in \citet{2018MNRAS.478.5091F}. The parameters of a photo-function \citep{1969A&A.....3..455M,2005MNRAS.362..542B} and 4$^{\rm th}$ order polynomial are fit and the relative magnitude $\displaystyle m^i_{\lambda, c}$ for each star $i$ is determined from the instrumental magnitudes $\displaystyle m^i_{\lambda, in}$ following Eq.~\ref{eq_cal1}, where $\lambda$ indicates the filter used.

\begin{equation}
m^i_{\lambda,c} = \mathcal{F} \left( m^i_{\lambda,in} \right) = A \cdot \log \left( 10^{B \left( m^i_{\lambda, in}-C \right )}+1 \right) + \mathcal{P}_4 \left( m^i_{\lambda, in} \right)
\label{eq_cal1}
\end{equation}

Due to the fact that a significant fraction of the amateur data are taken using slightly non-standard filters, or under non-photometric conditions, a second additional calibration step to correct colour terms needs to be applied to all data. The details of which are described in \citet{2020MNRAS.493..184E}. We first identify non-variable stars in all fields, by selecting objects with a low Stetson index $\mathcal{J}$ \citep{1996PASP..108..851S} and measure their magnitudes and colours. These stars are then used to fit the parameters of a second order polynomial in magnitude and colour (without cross-terms), including a common off-set $\displaystyle p_0$, to correct the colour terms. The final calibrated magnitudes $\displaystyle m^i_{\lambda,f}$ for each star are then determined following Eq.~\ref{eq_cal2}. For the purpose of this work we use the $\displaystyle V^i-I^i$ colour of the stars, but any other colour can be chosen. This colour is determined for each star from the median of the $\displaystyle m^i_{V,c}$ and $\displaystyle m^i_{I,c}$ values, that are available within five days of the image that is calibrated.

\begin{equation}
m^i_{\lambda, f} = \mathcal{W}(m^i_{\lambda, c}, V^i-I^i) = p_0 + \mathcal{P}^2(m^i_{\lambda, c}) + \mathcal{P}^2(V^i-I^i)
\label{eq_cal2}
\end{equation}

This colour correction usually significantly improves the root mean square (RMS) scatter of the magnitude offsets of all calibration stars. We use this scatter of stars which have the same magnitude (within $\pm 0.1$~mag) as the star in question, to determine the photometric uncertainty of the calibration \citep{2020MNRAS.493..184E}. All measurements with an uncertainty above 0.3~mag are excluded from the light curves analysed in this work.

\begin{figure*}
\centering
\includegraphics[angle=0,width=1.01\columnwidth]{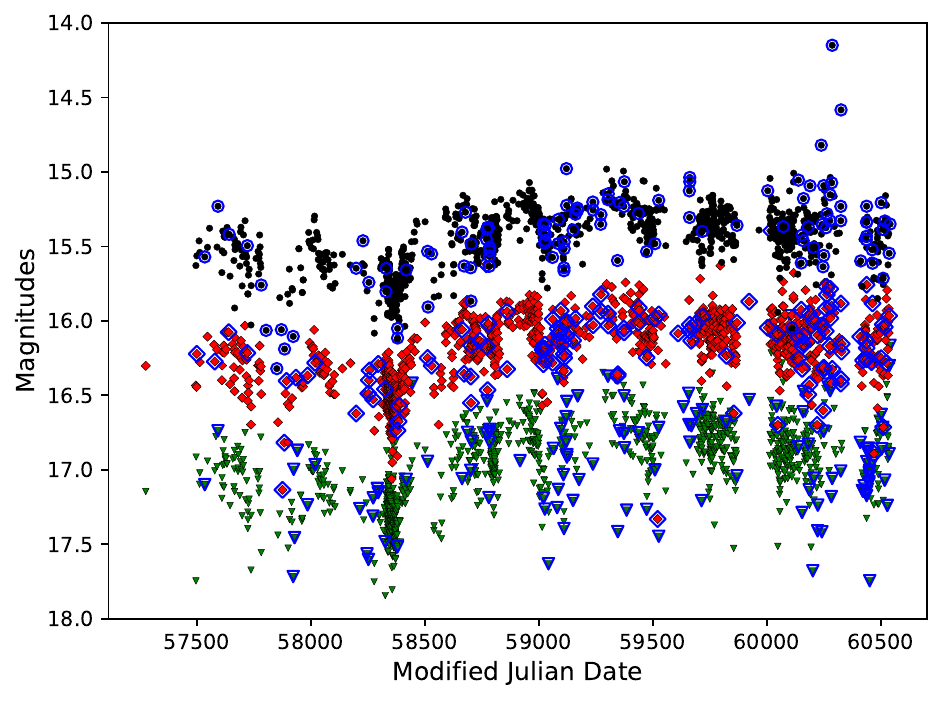} \hfill
\includegraphics[angle=0,width=0.99\columnwidth]{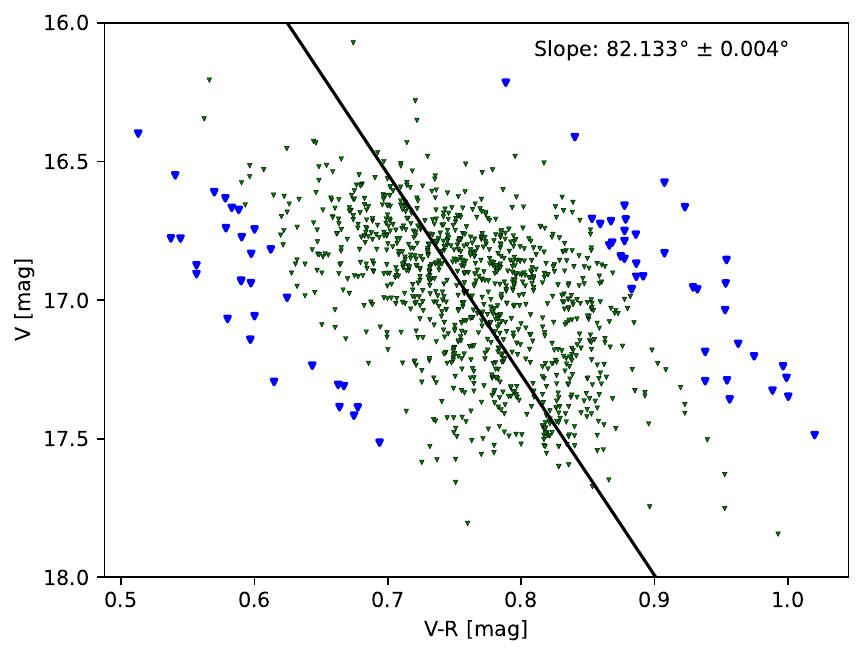} 
\caption{{\bf Left:} Example light curve of the variable source FHK~116. We show the $V$ (green triangles), $R$ (red diamonds), and $I$-band (black circles) light curves. The points highlighted in blue have been flagged up as potentially erroneous and have been removed from any subsequent analysis. {\bf Right:} Example $V-R$ vs $V$ colour magnitude plot for the same source. The best fitting slope is over plotted and the 3-sigma outliers are indicated in blue. Note the very different scales on the x and y-axis.} \label{lc_example}
\end{figure*}

\subsection{Removing Photometry Outliers}\label{cleaning}

The above detailed calibration ensures that the systematic photometry uncertainties are minimised. The nature of our HOYS data requires further data quality control to remove potentially erroneous data. Here, we detail the additional selection criteria applied to our data before they are being used for analysis.

i) Photometry closer than 5~arcmin to very bright stars ($Gmag < 6$~mag) has been removed for observatories where we have identified problems with PSF wings influencing the photometry (incl. the University of Kent’s Beacon observatory).

ii) Some stars apparently have two brightness measurements on the same image. This is either caused by good seeing and the object being a binary (apparent or real), or tracking issues in the image. All such photometry has been removed.

iii) We are not interested in very rare and short-term events such as flares. Thus, for each light curve and filter, we determine the mean and RMS of all magnitude measurements. Any points that deviate more than four times that value from the mean are removed.

iv) For each light curve a $V$ vs $R-I$ colour magnitude diagram (CMD) is created. For each $V$ magnitude measurement the $R-I$ colour is determined as the difference of the average of all $R$ and $I$ measurements within two days of the $V$ data being taken. A linear perpendicular distance fit to the CMD has been obtained, and the RMS scatter perpendicular to the line of best fit has been determined iteratively. All points further away from the line of best fit than three times that RMS value are removed from the light curve. This is repeated for all possible combinations of the $V$, $R$, and $I$ filters in the CMD, i.e. nine times.

In Fig.~\ref{lc_example} we show how the procedure performs for a typical variable source. In this case the object FHK~116 \citep{2022AJ....163..263H}, which is also known as 2MASS~J20513340+4434543 or Gaia~DR3~2163149219097677568. The left panel shows the $V$ (green triangles), $R$ (red diamonds), and $I$-band (black circles) light curves. We highlight in blue all data points that have been flagged up as potentially erroneous by our procedure and that are excluded from analysis. In the right panel of Fig.~\ref{lc_example} an example $V-R$ vs $V$ colour magnitude plot for the same source is displayed. The best linear fit and the outliers that are removed are indicated.

\begin{figure*}
\centering
\includegraphics[angle=0,width=0.33\textwidth]{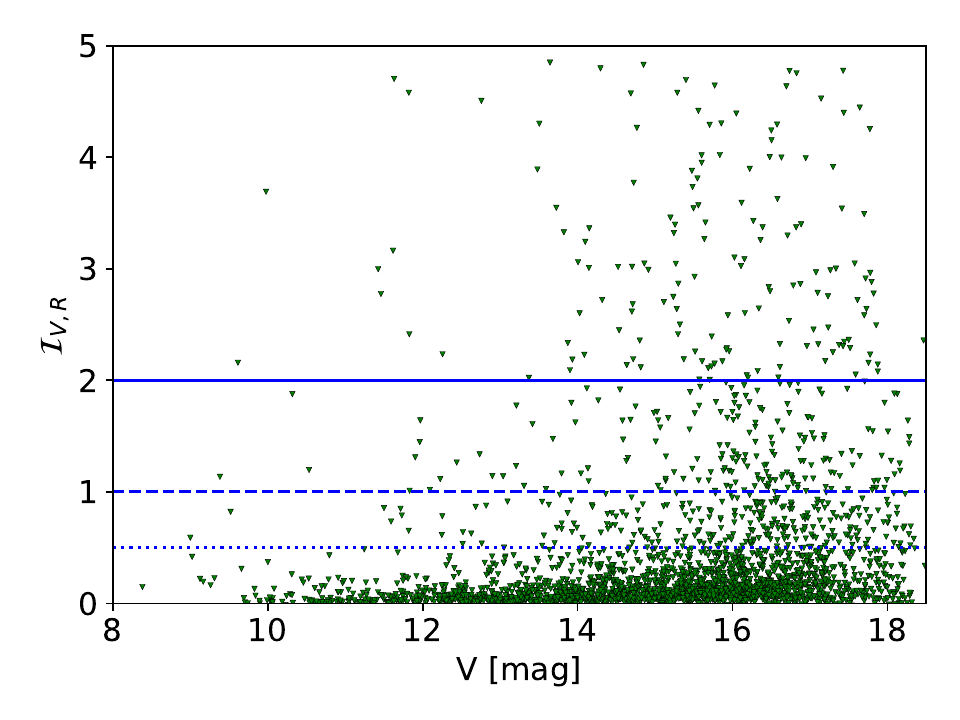}\hfill
\includegraphics[angle=0,width=0.33\textwidth]{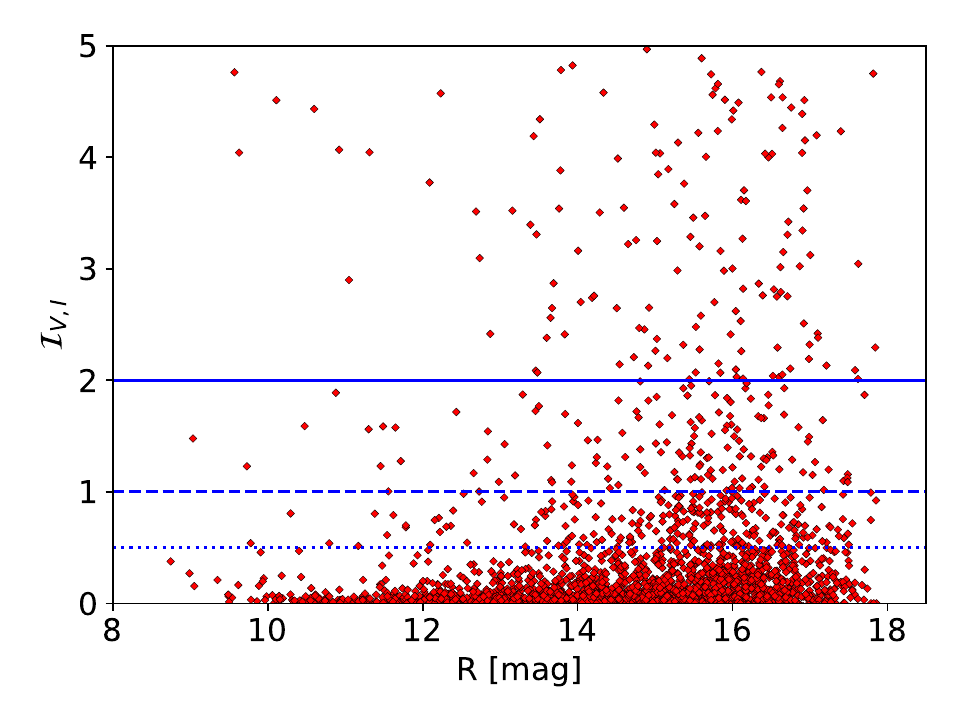}\hfill
\includegraphics[angle=0,width=0.33\textwidth]{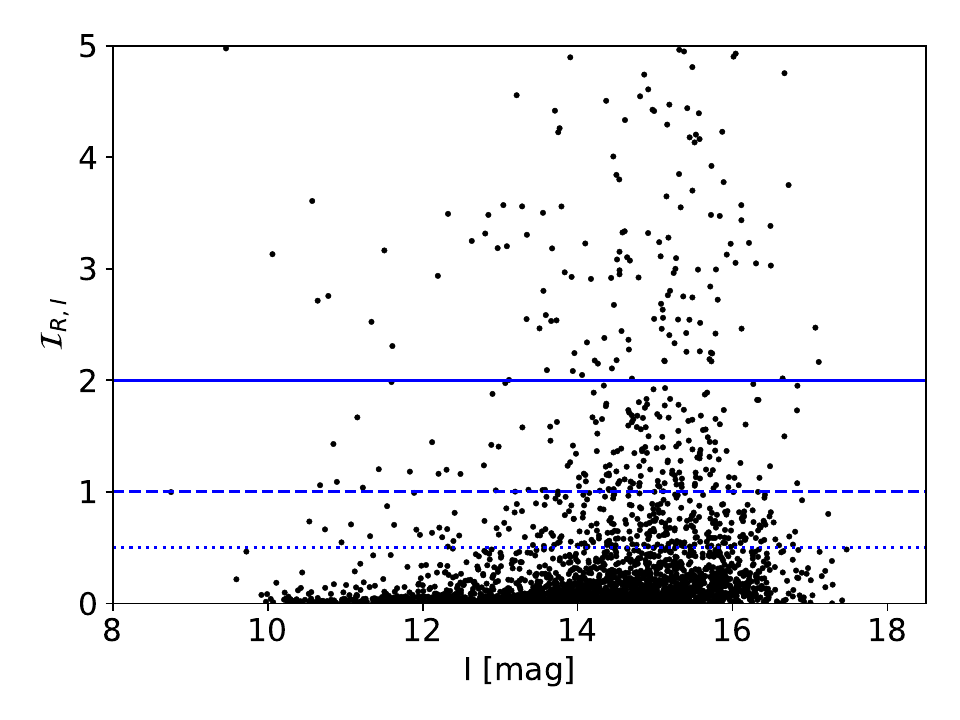}
\caption{Welch-Stetson index of our YSO sample determined from all three filter combinations, $\mathcal{I}_{V,R}$, $\mathcal{I}_{V,I}$, and $\mathcal{I}_{R,I}$ from left to right. The cut-off for highly variable objects used in this study is marked at $\mathcal{I} = 2$ with a solid line. We also indicate $\mathcal{I} = 1$ and $0.5$ with a dashed and dotted line, respectively. \label{stetson} }
\end{figure*}

\subsection{Identification of Variables}

The above data extraction (performed on June 18$^{th}$, 2024), calibration and 'cleaning' has been applied to the just over 3000 light curves of potential HOYS cluster members identified in \citet{2024MNRAS.529.1283F} based on their GAIA~DR3 astrometric properties (parallax and proper motion). This includes all clusters, even if they were not analysed in detail in that paper. No selection based on any other object properties (such as disk bearing or not) have been made. Here we only require a sample of clearly variable light curves to test our methodology (see Sect.~\ref{sect_clustering}). We removed all objects that had less than 100 data points in $V$, $R$, and $I$ remaining after the cleaning.

For each light curve, we determined the Welch-Stetson Index $\displaystyle \mathcal{I}$ \citep{1996PASP..108..851S}. This metric measures the correlation of the variations between two different filters, and is defined for the combination of $V$ and $I$ filters in Eq.~\ref{sw_ind}. It relies on having $N$ contemporary pairs of brightness measurements and their uncertainties in the two filters. These are determined in the same way as for the colours in the CMDs (see Sect.~\ref{cleaning}), over a time period of two days. 

\begin{equation}\label{sw_ind}
\mathcal{I}_{V,I} = \sqrt{\frac{1}{N(N-1)}} \sum_{i=1}^{N} \left( \frac{V_{i} - \overline{V}}{\sigma_{V,i}} \right) \left( \frac{I_{i} - \overline{I}}{\sigma_{I,i}} \right)
\end{equation}

In this work, we focus only on the data in the $V$, $R$, and $I$ filters. Thus, we have three possible combinations for the Welch-Stetson Index: $\displaystyle \mathcal{I}_{V,I}, \mathcal{I}_{V,R}$, and $\displaystyle \mathcal{I}_{R,I}$. In Fig.~\ref{stetson} we show how these depend on the apparent magnitude of the sources. The majority of stars shows only low level variability, but the $\displaystyle \mathcal{I}$ values go up to values of 1323. In our analysis, we found that $8, 5$, and $6$~\% of the $\mathcal{I}$ values were greater than five for $\mathcal{I}_{V,R}, \mathcal{I}_{V,I}$, and $\mathcal{I}_{R,I}$, respectively. These are not shown in Fig.~\ref{stetson} for clarity. For our subsequent analysis of highly variable sources, we select all stars that have an $\displaystyle \mathcal{I}$ value above two for all three filter combinations. Such values usually indicate significant variability above the level of observational noise \citep[e.g. ][]{2001AJ....121.3160C,2014AJ....147...82C}. This creates a sample of 240 highly variable sources. Note that the specific choices of time window lengths used to define contemporaneous brightness pairs for the Welch-Stetson Index, as well as the threshold for selecting highly variable sources, do not affect our subsequent results.

\section{Variability Fingerprints}\label{sect_fingerprints}

In this section, we describe how our YSO light curves are converted into variability fingerprints. We discuss their range, resolution, and how their uncertainties can be estimated.

\subsection{Fingerprint determination}

The light curves of all investigated variable YSOs have individual cadences. This is caused by the nature of HOYS, with its distributed observing. Thus, almost no two light curves can be directly compared, and we need to create a measure that allows us to capture the statistical properties of a star's variability and to compare this between objects. These are the variability fingerprints.

We have introduced these fingerprints generated for the HOYS data in \citet{2020MNRAS.493..184E}. They are based on earlier works, e.g. by \citet{2004A&A...419..249S,2015ApJ...798...89F,2017MNRAS.465.3889R}. The purpose of these variability fingerprints is to determine the probability $\displaystyle P(\Delta t, \Delta m)$, that a star varies by an amount $\displaystyle \Delta m$ when observed at a time difference $\displaystyle \Delta t$ apart. This can be done for light curves obtained in any filter. It assumes that the total length of the light curve and the observing cadence is sufficient so that all timescales and magnitude changes have been covered. This is obviously not the case for rare events, such as major bursts or long dimming events. Thus, events that only occasionally occur in a fraction of light curves can only be captured statistically, by analysing a large sample of light curves and their fingerprints.

The determination of the fingerprints for a light curve with $N$ data points ($t_i, m_i$: $i = 1 ... N$) is done by the following procedure:

i) We determine all $N \cdot (N-1) / 2$ pairs of positive time differences $\Delta t = t_j - t_i$ (with $t_j > t_i$) in the light curve, as well as their corresponding magnitude difference $\Delta m = m_j - m_i$.

ii) The distribution of the ($\Delta t, \Delta m$) values is converted into a two-dimensional histogram. There are hence $\mathcal{T}$ columns for the bins in $\Delta t$ and $\mathcal{M}$ rows for bins in $\Delta m$. The bin-sizes chosen are discussed below in Sect.~\ref{fp_range}. Each of the cells ({\it pixels}) in this histogram contains the number of ($\Delta t, \Delta m$) pairs, i.e. $N(x,y)$ with $x = 1 ... \mathcal{T}$ and $y = 1 ... \mathcal{M}$.

iii) To obtain the probabilities $\displaystyle P(\Delta t, \Delta m)$, the counts in each pixel are normalised by the total number of counts in the same column (range of $\Delta t$ values) they are in. This is shown in Eq.~\ref{hist_norm}. 

\begin{equation}\label{hist_norm}
P(x,y) = N(x,y) / \sum_{i=1}^\mathcal{M} N(x,i)
\end{equation}
 
\begin{figure*}
\centering
\includegraphics[angle=0,width=0.68\columnwidth]{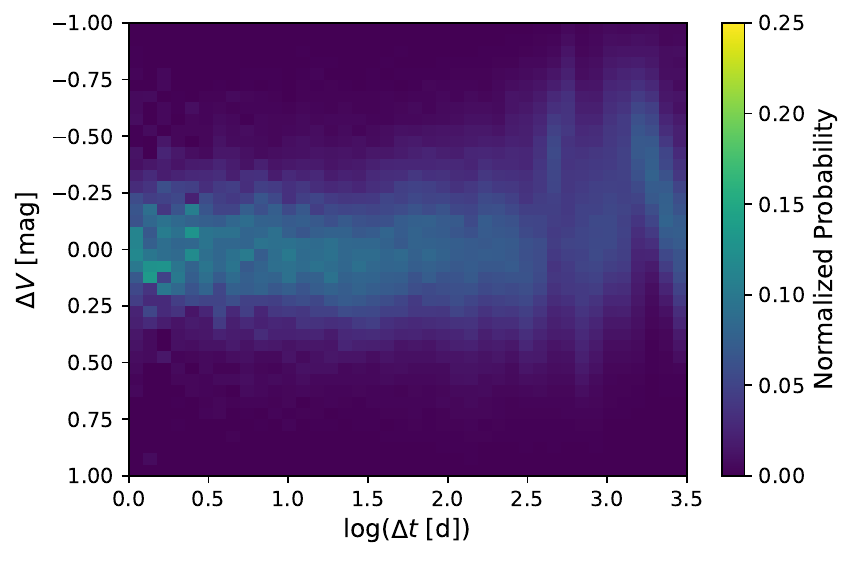} \hfill
\includegraphics[angle=0,width=0.68\columnwidth]{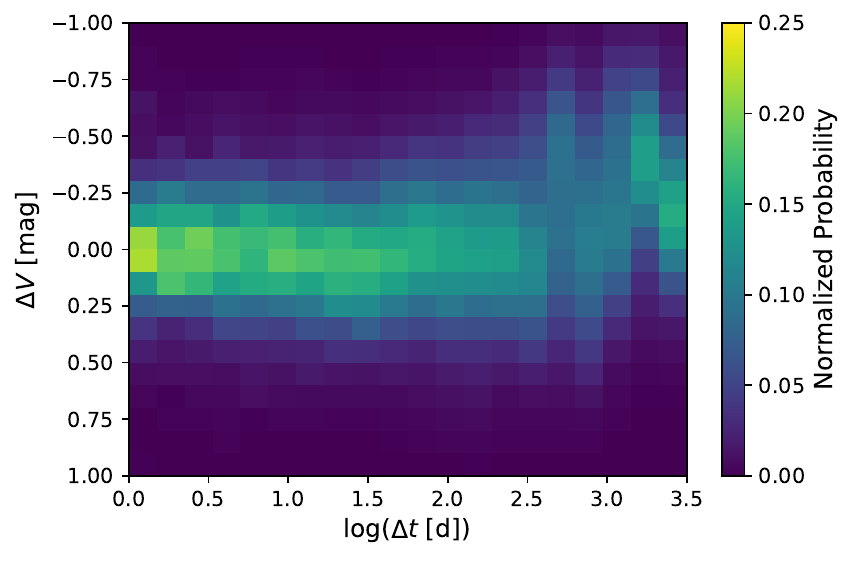} \hfill
\includegraphics[angle=0,width=0.68\columnwidth]{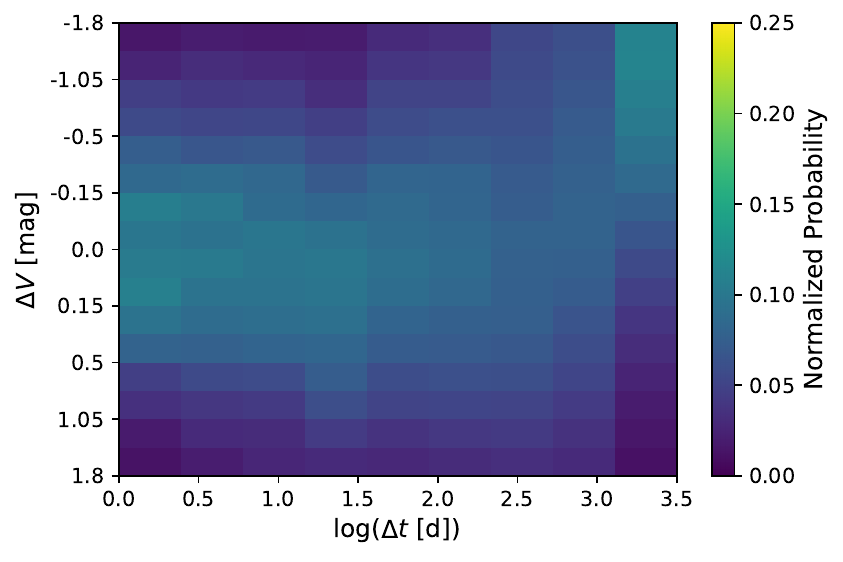}
\caption{Example $V$-band variability fingerprints of the source FHK~116. From left to right we show the resolutions 40~x~40 , 20~x~20, and 9~x~16 adaptive pixels. The range along the time axis are identical in all panels, but the adaptive pixel map on the right has a larger magnitude range than the other two maps. \label{fp_example} }
\end{figure*}

\subsection{Fingerprint Range and Resolution}\label{fp_range}

For our analyses, we need to choose the minimum and maximum ranges as well as the bin sizes and scaling in the variability fingerprints along both axes. In particular, the bin size determines the number of counts in each bin and thus the accuracy by which the probabilities can be determined. These will be discussed in Sect.~\ref{fp_accuracy}. Furthermore, the bin size sets the range of $\Delta t$ and $\Delta m$ over which the individual probabilities $P(\Delta t, \Delta m)$ are determined.

Along the $\Delta t$ axis the maximum range is set by our survey duration, which is up to 10~yr for some of the HOYS fields. Based on the design of the survey and our science goals, we set a minimum range of 1~d. It is not practical to use linearly scaled bins along the $\Delta t$ axis due to the wide range of scales. Thus, we use a $\log$ scaling (base 10) for the bin size in $\Delta t$, between the borders of $0.0 \le \log{(\Delta t [d])} \le 3.5$. This corresponds to time differences from one day to 8.67~yr. Equally sized bins in this log-scaling then correspond to the same ratio of time differences. For example, a bin size of $\log{(\Delta t [d])} = 1$ means that the bin contains time differences that are a factor of ten apart from smallest to largest, and bins of size $\log{(\Delta t [d])} = 0.3$ would correspond to time differences that are a factor of two apart from smallest to largest.

Along the $\Delta m$ axis, the range is set by the typical maximum variability that the investigated objects have. We find that for most of our variable objects, this reaches about $\pm 1-2$~mag. However, a few sources vary by more. Using linearly scaled bin sizes along the $\Delta m$ axis is a valid choice. The bin size should then not be below the typical photometric uncertainty of the brightness measurements, which is of the order of a few percent \citep[e.g.][]{2018MNRAS.478.5091F,2020MNRAS.493..184E}. Thus, we use a bin size of 0.05~mag. 

However, for larger variability amplitudes, the behaviour of a source is not expected to change significantly from one such bin to the next. In other words, the probability that a star varies by $+1.00$~mag on a given timescale should be very close to the probability that it varies by $+1.05$~mag on the same timescale. Hence, we are using {\it adaptive} bin sizes along the $\Delta m$ axis, where each row of pixels away from the $\Delta m = 0$~mag axis has the range $\Delta m$ increased by a value of $\mathcal{X}$ times the range of the first pixel. A value of $\mathcal{X} = 0$ corresponds to linearly scaled pixels discussed above. For all values of $\mathcal{X}$ between zero and a few, the results obtained do not significantly influence the results. We hence use $\mathcal{X} = 1$ as an example throughout the remainder of this paper. This means that the first pixels range from $0.00$~mag to $\pm 0.05$~mag. Each subsequent pixel is wider by the amount of $0.05$~mag. Thus, the second pixels range from $\pm 0.05$~mag to $\pm 0.15$~mag, the third pixels from $\pm 0.15$~mag to $\pm 0.30$~mag, etc.. This allows us to cover the variability range from $-1.8$~mag to $+1.8$~mag with only 16 of these adaptive pixels. The range and pixel number can be adjusted by choosing a different value for $\mathcal{X}$.

In Fig.~\ref{fp_example} we show several $V$-band fingerprints with different resolutions (including one with adaptive pixels) for the object FHK~116, the same source for which the light curves are plotted in Fig.~\ref{lc_example}. Up to time scales of about one year, the source does not vary considerably. On larger time scales, the object is more likely to brighten than to faint, which is caused by the short dimming event that the object underwent near MJD~=~58,400~d. The higher the resolution, the more detail is obviously visible. However, the accuracy of the map also decreases with higher resolution. We note that, other than the number of pixels without data in the fingerprints or low signal-to-noise probabilities (see Sect.\ref{fp_accuracy} below), the exact choice of bin size and range along both axes has no qualitative influence on the results of the analysis performed in Sect.~\ref{sect_clustering}. Only minor quantitative differences will occur if any of these values are changed, which have no influence on the results and conclusions of our work.

\begin{figure*}
\centering
\includegraphics[angle=0,width=0.68\columnwidth]{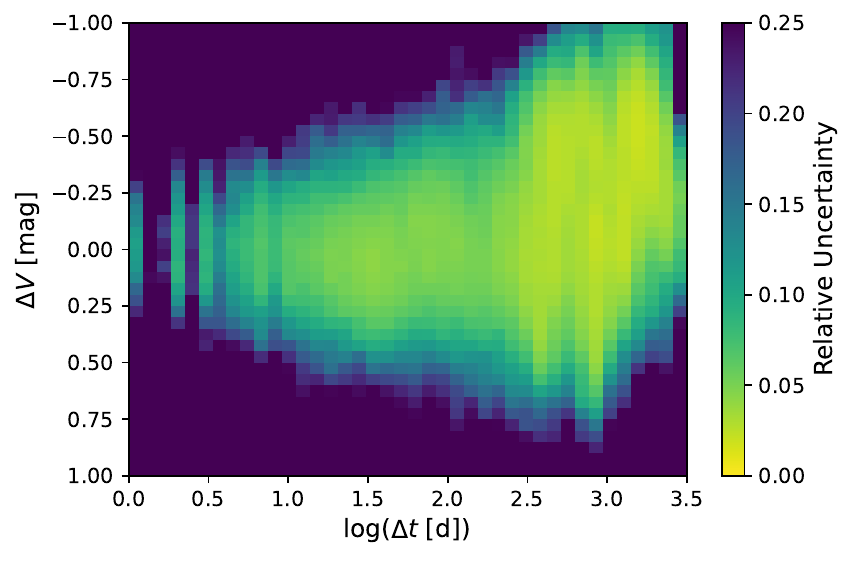} \hfill
\includegraphics[angle=0,width=0.68\columnwidth]{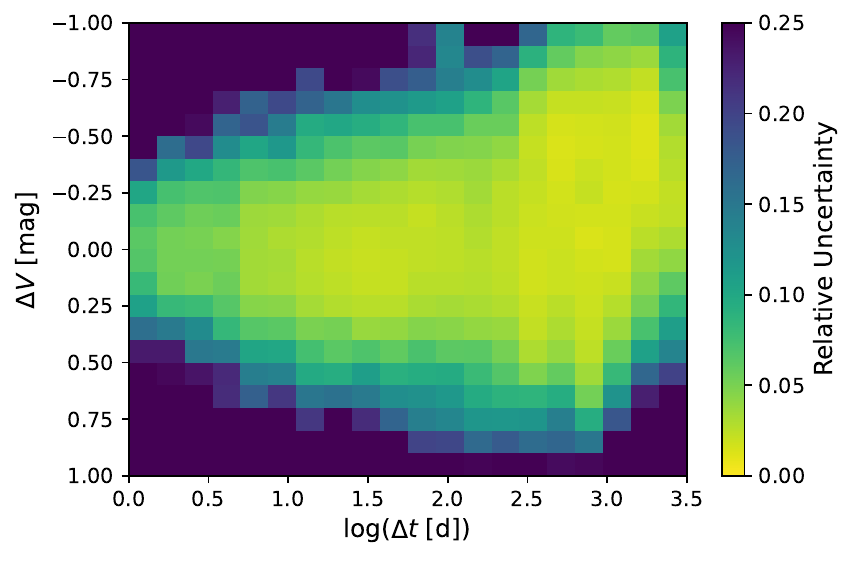} \hfill
\includegraphics[angle=0,width=0.68\columnwidth]{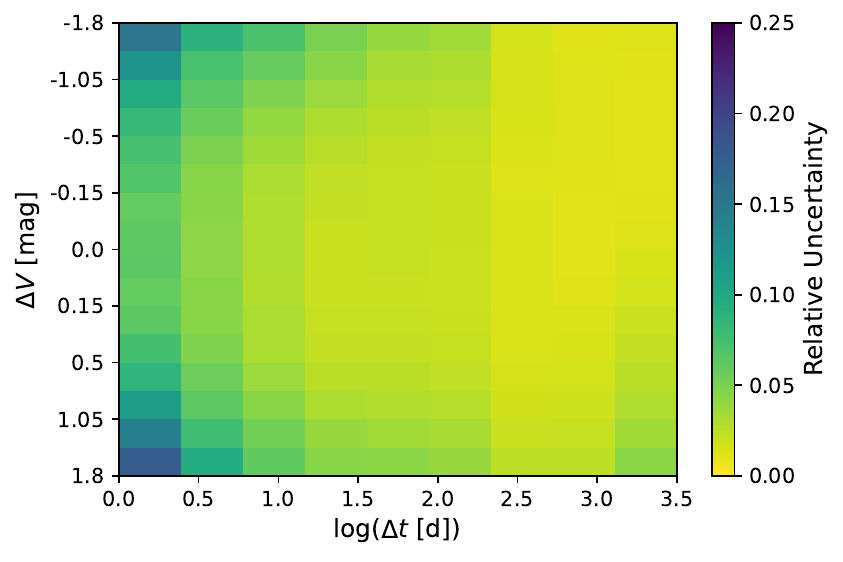} \\
\includegraphics[angle=0,width=0.68\columnwidth]{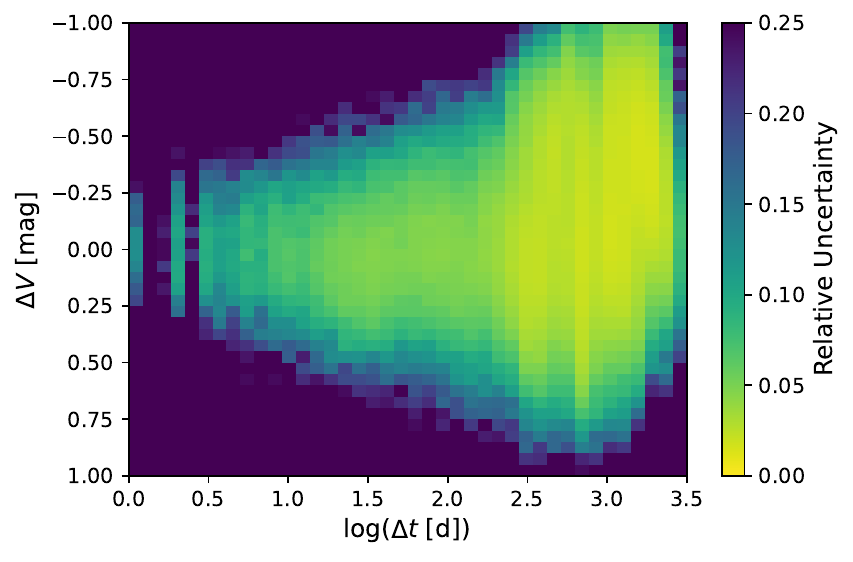} \hfill
\includegraphics[angle=0,width=0.68\columnwidth]{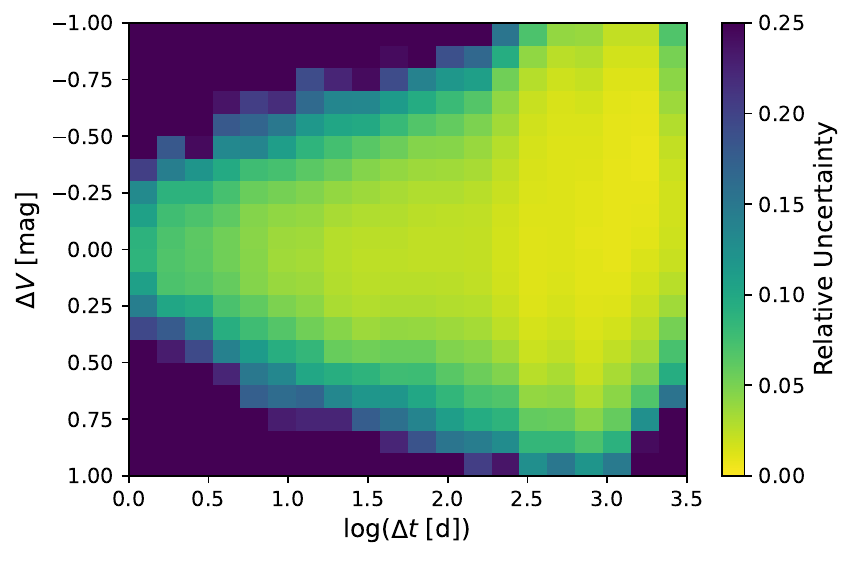} \hfill
\includegraphics[angle=0,width=0.68\columnwidth]{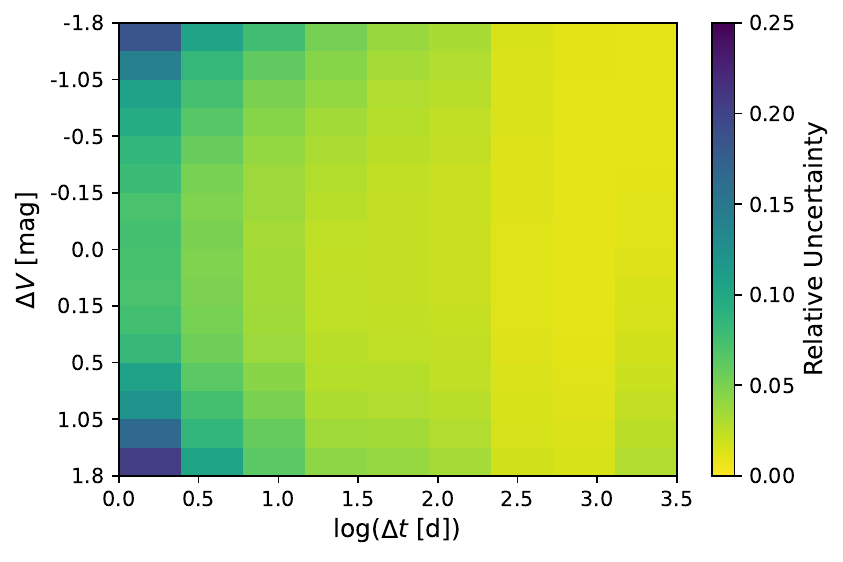} \\
\caption{\label{fp_errors}Example $V$-band variability fingerprint error maps of the source FHK~116. From left to right we show the resolutions 40~x~40 , 20~x~20, and 9~x~16 adaptive pixels. The top row shows the maps determined from bootstrapping the light curve, and the bottom row when using the Poisson counting statistics.}
\end{figure*}

\subsection{Fingerprint Accuracy}\label{fp_accuracy}

The probabilities $\displaystyle P(\Delta t, \Delta m)$ in the fingerprints have varying uncertainties, depending on the number of counts $\displaystyle N(\Delta t, \Delta m)$ in each pixel and how close the individual $(\Delta t,\Delta m)$ pairs are to the pixel boundaries. Ideally, we only want to include high significance $\displaystyle P(\Delta t, \Delta m)$ values in any subsequent analysis. Thus, we estimate the relative uncertainties of the fingerprints by two different methods: Light curve boot strapping and Poisson counting statistics.

\subsubsection{Light Curve Boot Strapping}

Each data point ($m_i$) in our light curve has an associated photometric uncertainty $\sigma (m_i)$ (see Sect.~\ref{data_calib}). We assume that there are no uncertainties for the times of the observation $t_i$ as they are in any case of the order of the exposure times (typically minutes), which is at least three orders of magnitude below the smallest time differences included in the fingerprints. 

To evaluate the uncertainties of the fingerprints, we repeat the determination of the fingerprint for boot strapped light curves. In other words, each light curve point is replaced by a random magnitude value. This value is chosen from a normal distribution with a mean of $m_i$ and a variance of $\sigma(m_i)$. This is repeated a number of times. We then use the average pixel value from all these fingerprints as the $\displaystyle P(\Delta t, \Delta m)$ value, and the standard deviation as its uncertainty $\displaystyle \Delta P(\Delta t, \Delta m)$. We have run extensive tests to evaluate how many of these boot strapping repeats are required to obtain accurate estimates for the fingerprint uncertainties. We find that typically after 30,000 repeats, the uncertainties will not vary by more than $10^{-4}$ from their value. Thus, this number was chosen for all boot strapping uncertainty calculations. 

In the top row of Fig.~\ref{fp_errors} we show some example $V$-band fingerprint relative uncertainty maps of object FHK~116 for several resolutions. It is evident that the accuracy decreases with increasing resolution. In particular for the highest resolution map (left panel), pixels that do not cover any integer day time difference have very high uncertainties due to our typical cadence of one day. Note that for the adaptive pixel maps (right panel), there are almost no low signal-to-noise pixels, even for the shortest time scales. 

\subsubsection{Poisson Counting Statistics}

The above described light curve boot strapping is very computationally intensive, since 30,000 fingerprints need to be calculated for each light curve to obtain the uncertainties. We have thus investigated if a much simpler (faster) method can lead to the same results. This is based on the assumption that the number of counts $N(x,y)$ in a pixel of the fingerprints follow the Poisson counting statistics. This would mean that their uncertainty is equal to the square root of their value. If this assumption is correct, the uncertainties for each pixel can be estimated by simple error propagation of Eq.~\ref{hist_norm}. This means the relative uncertainties of the fingerprints can be determined as shown in Eq.~\ref{err_hist}.

\begin{equation}\label{err_hist}
\frac{\Delta P(x,y)}{P(x,y)} = \sqrt{ \frac{1}{N(x,y)} + \left(\sum\limits_{i=1}^\mathcal{M} N(x,i)\right)^{-1} }
\end{equation}
 
In the bottom panel of Fig.~\ref{fp_errors} we show the relative uncertainty maps determined for the $V$-band fingerprints with the Poisson counting statistics, for the same maps as in the top row. We find that they are visually very similar, both qualitatively and quantitatively.

\subsubsection{Comparison of Fingerprint Uncertainties}

Here we briefly investigate the quantitative differences between the two uncertainty calculation methods. We have determined the uncertainty maps for both methods for all variable sources in our sample and for various resolutions. For each map a linear regression between the Poisson and boot strapping uncertainty values for all pixels has been performed. We have excluded all pixels in the maps where the relative uncertainties are worse than 0.33, i.e. where the signal-to-noise of the $\displaystyle P(\Delta t, \Delta m)$ values is below three. Note that such pixels are not very common in our maps (see e.g. Fig.~\ref{fp_errors}). In total only about 10 percent of pixels have such a low signal-to-noise, and these are exclusively confined to short timescales and high amplitudes. 

We find that the uncertainties based on the Poisson counting statistics are about $1.15 \pm 0.10$ times larger than the ones estimated from boot strapping the light curves. This factor does not systematically depend on the resolution of the fingerprints or the filter in which the light curve has been observed. Thus, we can conclude that the fingerprint relative uncertainties can be estimated using the much faster method that relies on the assumption of Poisson counting statistics. However, they should be scaled down by a factor of 1.15 to represent the correct relative uncertainties based on the light curve boot strapping. We note that if fingerprint pixel sizes are used that are significantly different to the ones in this work, the scaling factor should be re-evaluated.

\section{Comparing Variability Fingerprints}\label{sect_clustering}

As outlined above, we aim to find a methodology that allows us compare the YSO variability fingerprints and to potentially identify clusters amongst them. In future we plan to compare the observed fingerprints with fingerprints obtained from artificial observations of model disks. Thus, our goal is to establish a procedure that creates a preferably 2-dimensional representation of the distribution of fingerprints in a sample, which is also robust. In other words the addition of an individual object to the sample does not significantly change the distribution. We will refer to this kind of distribution as the fingerprint {\it landscape}, hereafter.

\subsection{Pre-analysis}\label{preanalysis}

The variability fingerprints are high dimensional data sets. Each pixel represents one dimension. For example, in the 9~x~16 adaptive pixel configuration we have 144 dimensions. Thus, to establish a 2-dimensional landscape (which can also be used for clustering) a dimension reduction tool needs to be applied. For the purpose of this paper we have tested Principle Component Analysis (PCA) and t-Distributed Stochastic Neighbour Embedding (t-SNE). We have also tested two clustering algorithms, DBSCAN and k-means. Both are iterative but have different approaches to selecting cluster membership. Centroids are used in k-means and core/border points in DBSCAN. The latter also uses t-SNE as part of its workflow.

We applied PCA using the standard scaler, which normalises the value in each dimension (fingerprint pixel) to a mean of zero and a standard deviation of one \citep{hastie2009elements}. This is sufficient as all pixel values in the fingerprints represent probabilities, i.e. are already limited to in-between zero and one. The dimension reduction was done for several configurations (9~x~16, 20~x~20, etc.) of our 240 fingerprints (obtained from $V$, $R$, or $I$-band data) to investigate how much of the variance can be captured in just two dimensions. Generally the configuration leading to the maximum value of variance captured, should lead to the best clustering results. 

We find that for all configurations, the first two components capture just shy of 50 percent of the total variance. As there is no significant difference between these values, all subsequent analysis is only presented for the 9~x~16 adaptive pixel configuration and $V$-band data. Note that the results for all other fingerprint resolutions and filters do not significantly differ qualitatively and quantitatively. Generally, much higher values than 50 percent for the total variance in the first two components are expected for distinct clusters \citep[e.g.][]{MurtyDevi}. Our low values thus indicate that there are most likely no distinct clusters identifiable in the fingerprint landscape, but that they rather form a continuum or single cluster.

\begin{figure*}
\centering
\includegraphics[angle=0,width=\columnwidth]{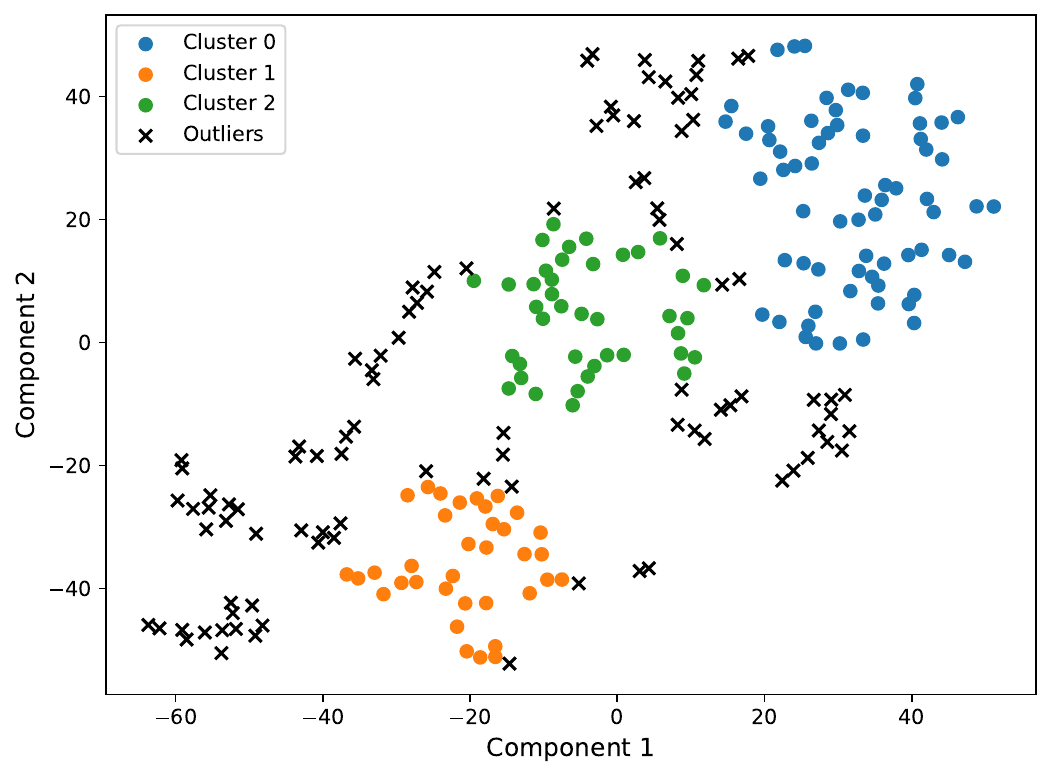} \hfill
\includegraphics[angle=0,width=\columnwidth]{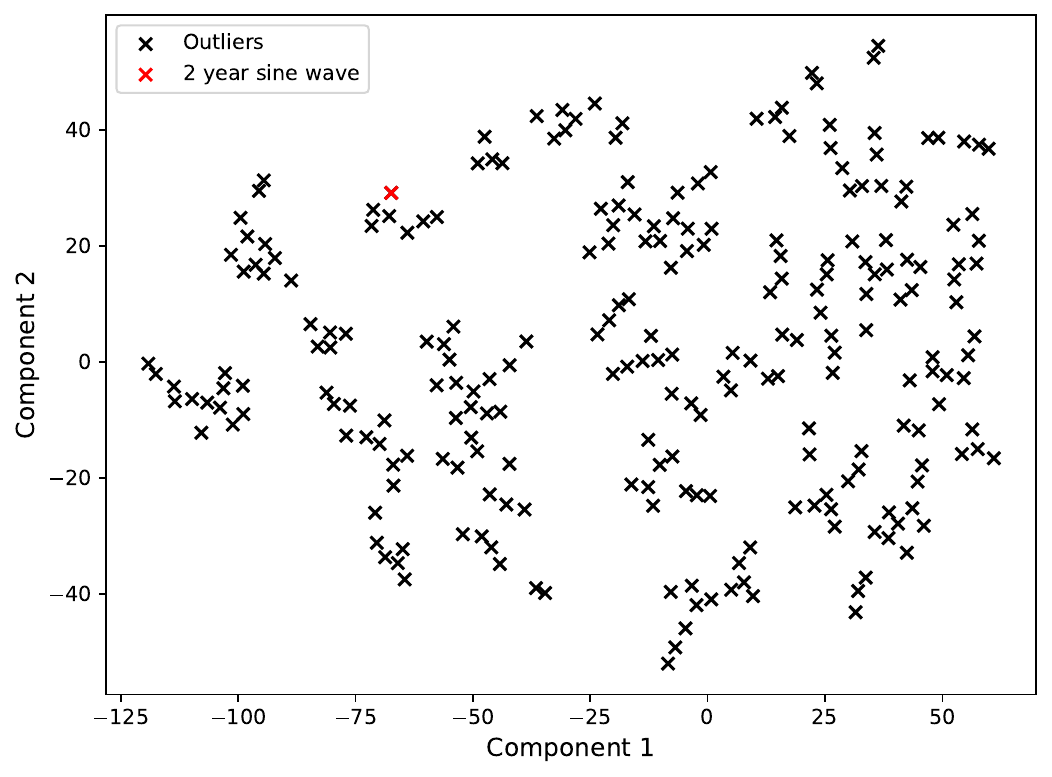}
\caption{\label{dbscan_results} {\bf Left:} DBSCAN clustering results for the 9~x~16 $V$-band adaptive pixel fingerprints of 240 variable YSOs. {\bf Right:} The same data and process as in the left panel but with one model fingerprint added which corresponds to a sine wave like light curve with an amplitude of 1~mag and a period of 2~yr.}
\end{figure*}

For DBSCAN three parameters must be determined. These are: epsilon ($\epsilon$), which is directly a part of DBSCAN, perplexity ($\mathcal{PP}$) which is a parameter for t-SNE, and the minimum number ($\mathcal{N}$) of objects in a cluster. The values for these parameters can be determined by calculating the Davies-Bouldin index (DBI) for their values. We used a range of 1~$\le \epsilon \le$~20, 5~$\le \mathcal{PP} \le$~50, and 5~$\le \mathcal{N} \le$~50 with the condition that the number of clusters is greater than one. These ranges do encompass typical accepted values \citep[$\mathcal{PP}$:][]{maaten2008tsne}, or are sensible given our data \citep[$\epsilon, \mathcal{N}$:][]{davies1979cluster, ester1996dbscan}. The combination that leads to the minimum DBI is the optimal value for the parameters. The minimum DBI obtained was approximately 0.45. The same value was obtained for a number of combinations of parameters, one of which was: $\epsilon$~=~16, $\mathcal{PP}$~=~7, and $\mathcal{N}$~=~35. The DBSCAN clustering result for this example is shown in the left hand panel of Fig.~\ref{dbscan_results}.

A DBI of under 0.40 suggest highly compact clusters that are well-separated \citep{davies1979cluster}. Our index does not meet this criterium but is close to it. However, there is a large number of outliers evident in Fig.~\ref{dbscan_results}, roughly 40~$\%$. This adds some further insight into the clustering outcome. The DBI index is a metric of the clusters formed only. The large number of outliers spread evenly throughout the left panel of the figure shows that there is no real separation between the clusters and the data as a whole are a continuum. For some of the combinations of parameters the DBI index was greater than two, indicating clusters that are either not compact, poorly separated, or both. Some analysis was also carried out on the parameter space using a Silhouette score. A maximum silhouette score of 0.4 was found for $\epsilon$~=~18, $\mathcal{PP}$~=~5, and $\mathcal{N}$~=~20. Generally, a silhouette score above 0.5 is required for separated distinct clusters \citep{MurtyDevi}. The low silhouette score could derive from the fingerprints forming a continuum or that it is not suitable as a metric for concave clusters such as those formed by DBSCAN \citep{rousseeuw1987silhouettes}.

Hence, all the indications show that the variability fingerprints in our sample do not form distinct clusters. They rather form a continuum distribution in the landscape. Our subsequent analysis hence focuses on the stability and robustness of this landscape. Hence, the clustering algorithms applied to the data in the subsequent sections are for illustrative purpose only. The above analysis clearly shows that no algorithm will identify real clusters in our data.

\subsection{Analysis with PCA, t-SNE and DBSCAN}

\subsubsection{Clustering with DBSCAN}

DBSCAN was chosen to illustrate clustering in our dataset as t-SNE can be made part of its workflow. In addition it is not sensitive to noise and will still form clusters. If there are outliers present they will be identified as such and not given cluster membership based on an arbitrary condition, e.g. the number of clusters set manually (as in k-means - see below).

We used the 240 object strong 9~x~16 $V$-band fingerprint sample and applied a dimension reduction using PCA from 144 down to ten components. The output was passed to t-SNE and the dimension reduced further to two. Note that changing the value from which t-SNE reduces the dimension to two has no influence on the results, as long as t-SNE is used at all. DBSCAN was used with $\epsilon$~=~16, $\mathcal{PP}$~=~7, $\mathcal{N}$~=~35, and 4,000 iterations (see Sect.~\ref{preanalysis}). The results of this clustering are shown in the left panel of Fig.~\ref{dbscan_results}. One can see that the fingerprints are grouped in 3 distinct clusters, but there is also a large number (102, or 43 percent) that have been labelled as outliers. We have visually inspected a number of randomly chosen pairs of light curves that have been assigned to the same cluster. They appear to show similar behaviour (see also Fig.~\ref{example_lcs} in the Appendix).

\subsubsection{Stability of PCA and t-SNE}

As indicated above we require stability of the fingerprint landscape. To analyse this we randomly selected one of the YSO light curves. The observing times (cadence) was maintained but the magnitude values were replaced by a sine wave with a 2~yr period and 1~mag amplitude. This was fingerprinted in the same way as all light curves and the fingerprint added into the sample. PCA and t-SNE were used for dimension reduction. DBSCAN was used to illustrate the clustering with exactly the same parameters as in the previous section, just with this one artificial object added. The results are shown in the right hand panel of Fig.~\ref{dbscan_results}.

There are significant differences in the landscape between the two versions. Some of the substructures are recognisable in both versions, but generally the entire structure has changed. However, points in close proximity do roughly remain in close proximity to each other. Notably, after adding just this one additional source, no clusters are formed at all, and all objects are classed as outliers. Thus, these significant changes in the landscape as seen in the two panels in Fig.~\ref{dbscan_results} render the method not useful for our purpose. We note that for other values of $\epsilon$, $\mathcal{PP}$, and $\mathcal{N}$ the changes can be slightly less severe, however there is no combination of these parameters that creates a stable landscape when one object is added.

\begin{figure*}
\centering
\includegraphics[angle=0,width=\columnwidth]{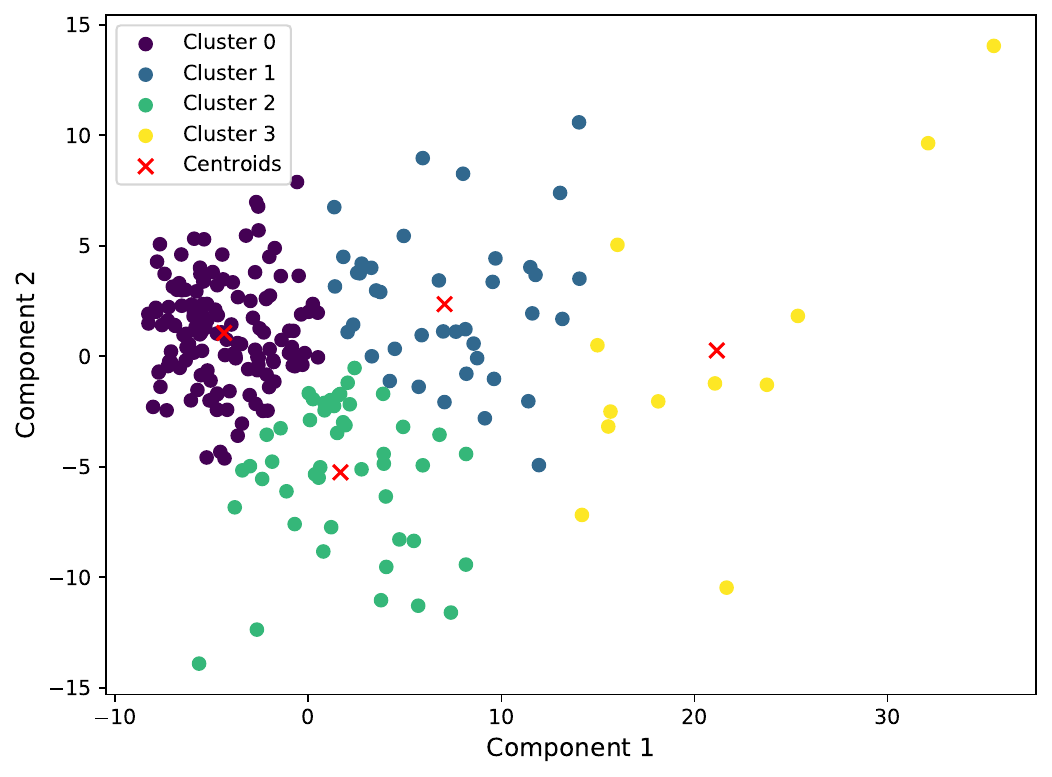} \hfill
\includegraphics[angle=0,width=\columnwidth]{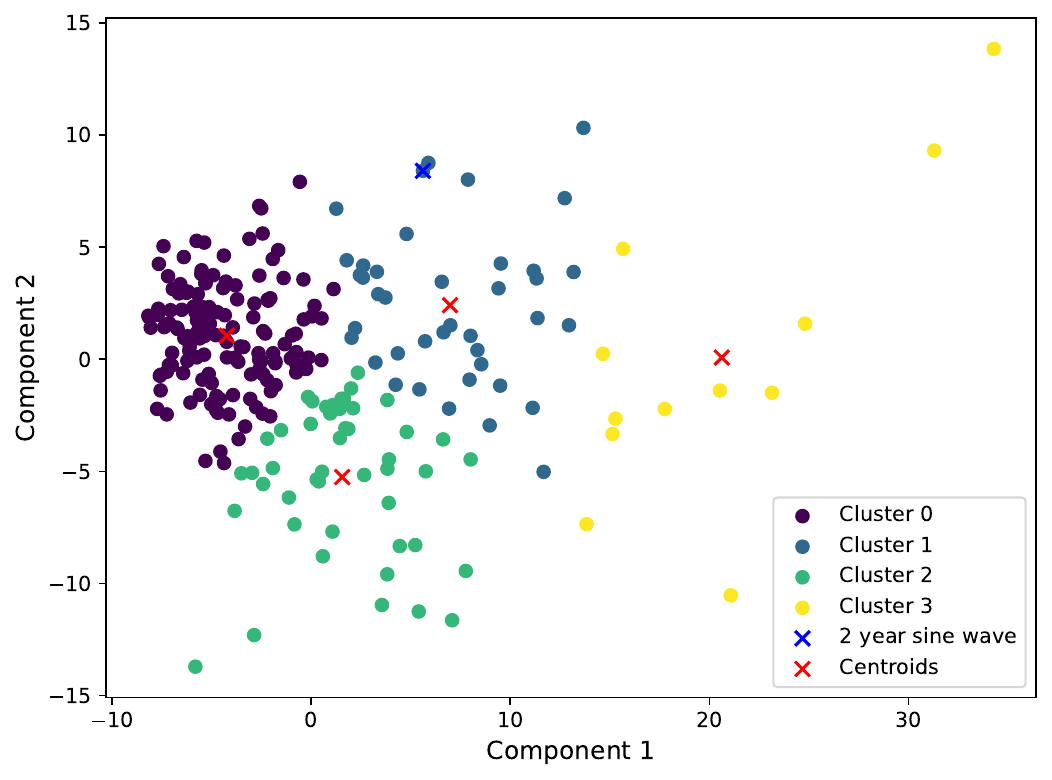}
\caption{\label{kmeans_results} {\bf Left:} Clustering results with PCA and k-means using the 240 sized sample the 9~x~16 $V$-band adaptive pixel fingerprints and 4 clusters. {\bf Right:} The same process as the left figure with one artificial fingerprint added which corresponds to a sine wave like light curve with an amplitude of 1~mag and a period of two years.}
\end{figure*}

We have also done extensive tests using different randomly selected light curves to create the artificial fingerprint, different fingerprint resolutions, filters of the photometry ($V$, $R$, $I$), and DBSCAN parameters ($\epsilon$, $\mathcal{PP}$, $\mathcal{N}$). In all cases we obtain the same qualitative and quantitive results. This method does not generate a stable landscape of points. This is caused by the non-linearity of the t-SNE process and the fact that our data are not clustered but form a continuum, as indicated by the DBI scores obtained (see Sect.~\ref{preanalysis}). Thus, while the combination of PCA, t-SNE, and DBSCAN has some utility in grouping light curves with similar properties into clusters the instabilities in the generated landscape caused by t-SNE makes it unsuitable for our aim to insert an artificial light curve for comparison with the observed YSO sample.

\subsection{Analysis with PCA and k-means}

\subsubsection{Clustering with k-means}

The k-means clustering is generally thought of as a less sophisticated algorithm than DBSCAN. It converges quickly often within a few iterations and has only one parameter (the number of clusters) which needs to be set. This determines the number of centroids around which the clusters are formed. The process is sensitive to outliers and noisy data. Here we illustrate the k-means analysis of our data for the 9~x~16 adaptive pixel fingerprints made from the $V$-band light curves of 240 variable YSOs - the same as used in the previous section. We used 4 clusters as the input for k-means, but note that specific results obtained do not depend on the exact value chosen. Scaling was applied using the standard scaler and PCA was applied to reduce the dimensions from 144 to two. 

The visual representation of the results of using PCA and k-means to illustrate the clustering can be seen in the left hand panel of Fig.~\ref{kmeans_results}. Unlike the DBSCAN results in Fig.~\ref{dbscan_results}, the clusters formed by k-means have less separation. The landscape created by PCA is clearly a continuum instead of distinct clusters. A visual inspection of light curves assigned to different clusters but close to each other in the landscape shows that they subjectively look more similar than light curves of objects in the same cluster but far away from each other in the landscape. This is further illustrated by Fig.~\ref{example_lcs} in the Appendix and also holds for artificial light curves (see later in Sect.~\ref{pca_artificial}).

The differing outcomes observed when applying DBSCAN versus k-means are not a result of inherent structures in the data (the fingerprints represent a continuum distribution - see Sect.~\ref{preanalysis}), but rather stem from the properties of the dimensionality reduction and clustering methods themselves. K-means, typically applied after linear transformations such as PCA, is limited to identifying convex clusters centered around means. In contrast, DBSCAN, when used following a non-linear transformation like t-SNE, can detect arbitrarily shaped clusters due to t-SNE’s tendency to exaggerate local structure \citep{Wangetal}. However, this enhanced local separation does not indicate true clustering, especially in the absence of any real discrete groups. Instead, the apparent clusters produced by DBSCAN after t-SNE, especially with low perplexity and appropriate density parameters, are artifacts of these methods imposing structure onto fundamentally unclustered data. This highlights how combinations of dimensionality reduction and clustering techniques can generate misleading results when applied to continuous datasets.

\subsubsection{Stability of PCA}\label{txt_kmeans_stability}

We have established that the fingerprint landscape does not exhibit distinct clusters, and that the combination of PCA and t-SNE does not yield a stable configuration. We now assess whether using PCA alone produces a stable landscape. As in the earlier DBSCAN test, one light curve was randomly selected, and its original cadence retained while replacing its magnitudes with a synthetic sine wave (period: 2 yr, amplitude: 1 mag). A fingerprint was generated for this artificial light curve and added to the dataset before applying PCA. The resulting landscape is shown in the right-hand panel of Fig.~\ref{kmeans_results}, with k-means applied solely to illustrate the spatial distribution of fingerprints.

A comparison of the two panels in Fig.~\ref{kmeans_results} shows that the positions of the original 240 fingerprints do not change by more than a small fraction of the typical separation between points (this will be discussed in Sect.~\ref{pca_artificial}). Thus, the association of fingerprints with the four illustrative clusters has remained almost unchanged. Only one object shifted from cluster 0 to cluster 1. The synthetic sine wave was associated with cluster 1. Crucially, the structure of the landscape itself remains stable after the inclusion of the additional object.

\begin{figure}
\centering
\includegraphics[angle=0,width=\columnwidth]{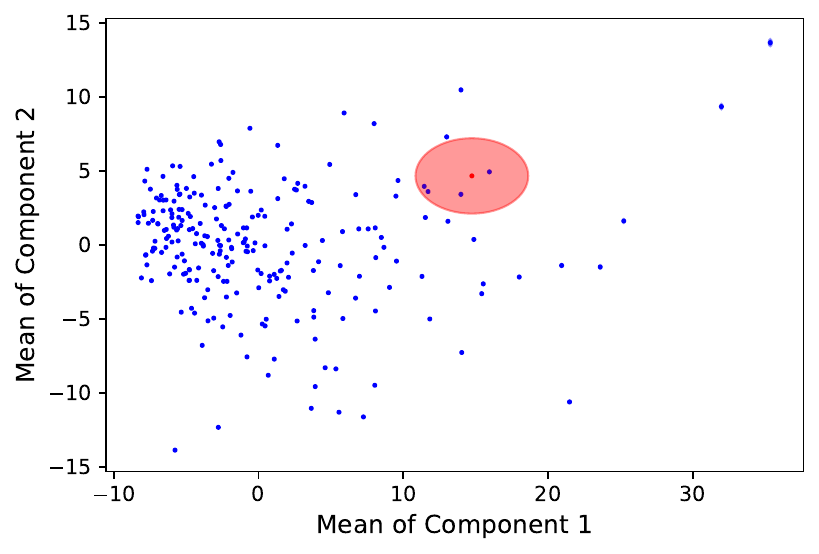} 
\caption{\label{kmeans_stability} Stability test using PCA for dimension reduction. A randomly selected object (highlighted in red) has its light curve bootstrapped 1000 times. The mean positions (points) and standard deviations (shaded ellipses) in the fingerprint landscape are shown where the standard deviations are scaled up by a factor of 20 for better visibility. }
\end{figure}

To further quantify this stability, we performed several additional tests. We only display  the landscape as determined by PCA in the subsequent figures, as we have established in Sect. 4.1, that there is no evidence of intrinsic clustering in our data. We utilise our 240 fingerprint sample and select randomly one of the objects. All individual photometry data points in that light curve are then bootstrapped, i.e. varied according to their photometric uncertainty. This means each magnitude is replaced by a number randomly drawn from a normal distribution with a mean value corresponding to the measured magnitude and a variance corresponding to its uncertainty. We created 1000 bootstrapped light curves (and fingerprints) for this object and determined the landscape with PCA for each of them separately with the other 239 objects. The results of the stability test for a typical source are shown in Fig.~\ref{kmeans_stability}. Note that the size of the uncertainty ellipse is governed by photometric noise of the object and at most objects have twice the variation of the one shown.

In that figure we show the average position in the landscape of the 1000 repeats for the unchanged objects as blue dots. The mean position for the bootstrapped light curve is shown in red. We also over plot as the diameter of the light blue and light red shaded ellipses the one sigma deviations from the mean for all objects. Note, that these deviations are scaled up by a factor of 20 for better visibility. However, the changes in the position within the landscape for the unaltered objects are still smaller than the symbol size used for the mean positions. Considering the scaling in Fig.~\ref{kmeans_stability}, we can conclude that the photometric uncertainty in the HOYS light curves does not significantly change the fingerprint landscape. At most the uncertainties in the photometry move positions in the landscape by less than the typical separation of nearby neighbours. 

\subsection{PCA with artificial light curves}\label{pca_artificial}

\begin{figure}
\centering
\includegraphics[angle=0,width=\columnwidth]{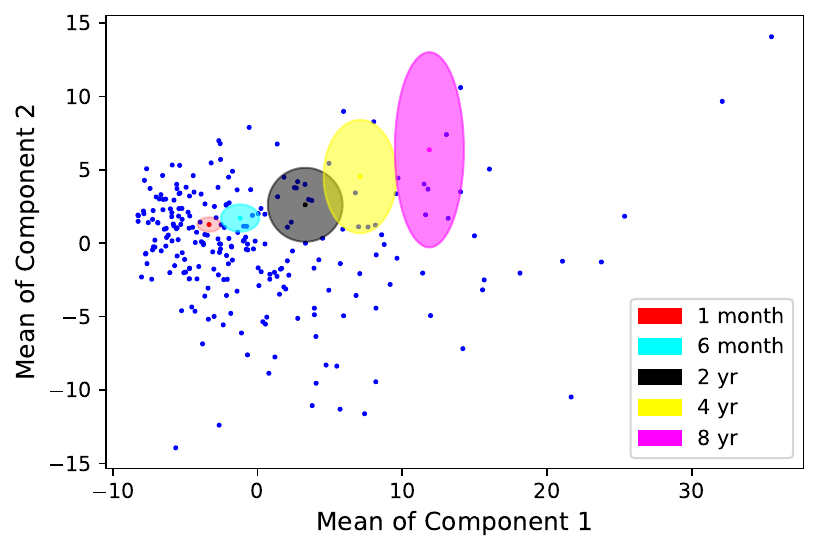}
\caption{\label{kmeans_sinewave} Stability test using PCA for our sample of 240 variable objects plus light curves representing sine waves. The approximate periods (see text for details) are indicated in the legend. The amplitudes of the sine waves are 1~mag in all cases. For each period 960 repeats with bootstrapped photometry, random phase shifts, and varying cadences are used to determine the mean positions (points) and their standard deviations (shaded ellipses). }
\end{figure}

We have established above that the fingerprints of light curves of variable young stars form a continuum distribution. However, in Sect.~\ref{txt_kmeans_stability} we have shown that the fingerprint landscape is stable when applying PCA for dimension reduction. Furthermore, the photometric uncertainties in the HOYS data have no significant influence on the position of an object in the landscape. As outlined in Sect.~\ref{intro} our long-term aim is to use artificial light curves from numerical models to investigate whether they are consistent with the observed fingerprints. Thus, we need to consider what influence the timing and observing cadence have on the position of a particular underlying light curve in the fingerprint landscape. We thus slightly change the stability analysis done above using two sets of artificial light curves.

\subsubsection{Sine Wave light curves}

We insert 2~yr period and 1~mag amplitude sine waves as artificial light curves (fingerprints) into the sample. This was done 960 times, and each of the 240 cadences of the original sample was used four times. In each case, a random phase shift was applied to the sine wave. This test shows how the observing cadence and the timing of the observations (phase shift for the sine wave like light curves) influence the position of the associated fingerprint in the landscape. We show the results in Fig.~\ref{kmeans_sinewave}. The mean positions for the 240 variable objects are shown as blue points. Their standard deviation is again too small to be plotted. The black point indicates the mean position of all 960 artificial sine waves with a 2~yr period and 1~mag amplitude. The diameter of the gray shaded ellipse represents the standard deviation from the mean, i.e. the typical scatter of the position of this type of light curve in the fingerprint landscape.

In contrast to the initial stability test, the standard deviations are much larger. This indicates that the timing of the observations and the observing cadence have an influence on the position of an artificial light curve (fingerprint) in the landscape. However, this range of positions is significantly smaller than the overall spread of points. This will hence allow us to evaluate qualitatively and quantitatively how much overlap there is in the variability characteristic of an artificial/model light curve and those of an observed sample of variable stars. Qualitatively, we can investigate where in relation to the observed fingerprints the model light curves are placed. If they are separated then the model leads to variability characteristics that are not present in the observed sample. If there is overlap, one can investigate whether there are characteristics in the light curve or YSO properties that distinguish objects inside and outside the overlap region in the landscape. A detailed discussion of this, including how the overlap of two samples can be quantified, how much overlap there is between YSO samples of different properties (e.g. evolutionary stages, inner disk excess emission, etc.), or model light curves, will be subject of a follow-up investigation and is outside the scope of this current work. Here, we solely focus on establishing and testing the methodology to create a stable fingerprint landscape based on observed light curves of variable YSOs.

We have performed additional tests identical to the one discussed above, with different sine wave like model light curves. We used five different periods for sine waves with amplitudes of 0.5 and 1.0~mag. The periods correspond roughly to one month, six months and 2, 4, and 8~yr. The exact values chosen (listed in Table~\ref{per_table}) ensure that the periods are not exact integer days to avoid sampling issues with our typical observing cadences. Obviously, typical YSO light curves are not represented by sine waves. The exceptions are spotted stars, but these generally have small amplitudes and short periods \citep[e.g.][]{2021MNRAS.506.5989F, 2023MNRAS.520.5433H}. Thus, one would expect only a small amount of overlap between these simple model light curves and the observed sample. 

The results of this test for the 1.0~mag amplitude sine waves are summarised in Fig.~\ref{kmeans_sinewave}. We can see that, regardless of the period and amplitude, the range in the landscape covered by the model fingerprints is small, compared to the spread of the entire YSO sample. The scatter in the landscape increases systematically with the period. This is understandable since our fingerprints only cover timescales of up to 8.5~yr. Thus, the shorter periods fit into all the light curves multiple times. Hence, their variability characteristics are fully sampled in each light curve. However, for longer periods each light curve only samples a part of the variations, thus causing the increased scatter. Furthermore, there are systematic changes of the average positions of the models with respect to the landscape. In particular, the placement is ordered by period, i.e. the closer the periods are to each other, the closer their placement in the landscape. Longer period light curves are placed systematically further away from the majority of observed fingerprints, to the right (positive values) along principle component~1 (PC~1). They are also systematically shifted towards slightly higher (positive) values along principle component~2 (PC~2). The same qualitative and quantitative behaviour is evident if we use a 0.5~mag amplitude in the sine waves. Below (Sect.~\ref{placement}) we investigate which light curve properties are responsible for the placement in the fingerprint landscape and discuss how this explains the systematic behaviour seen in the simulated sine wave like light curves.

\begin{table}
\caption{\label{per_table} Approximate periods (listed in Fig.~\ref{kmeans_sinewave}) and the exact irrational periods used in the simulated sine wave light curves.}
\centering
\setlength{\tabcolsep}{3.5pt}
\renewcommand{\arraystretch}{1.0}
\begin{tabular} {lccccc}
\hline 
Approx. Periods & 1~months & 6~months & 2~yr & 4~yr & 8~y \\[+0.5ex] \hline \\[-1.7ex]
Exact Periods & $\frac{1}{3.8 \pi}$~yr & $\frac{1}{0.66 \pi}$~yr & $\frac{1}{0.15 \pi}$~yr & $\frac{1}{0.08 \pi}$~yr & $\frac{1}{0.04 \pi}$~yr \\[+1ex] \hline
\end{tabular}
\end{table}

\subsubsection{Dipper and Burster light curves}

\begin{figure}
\centering
\includegraphics[angle=0,width=\columnwidth]{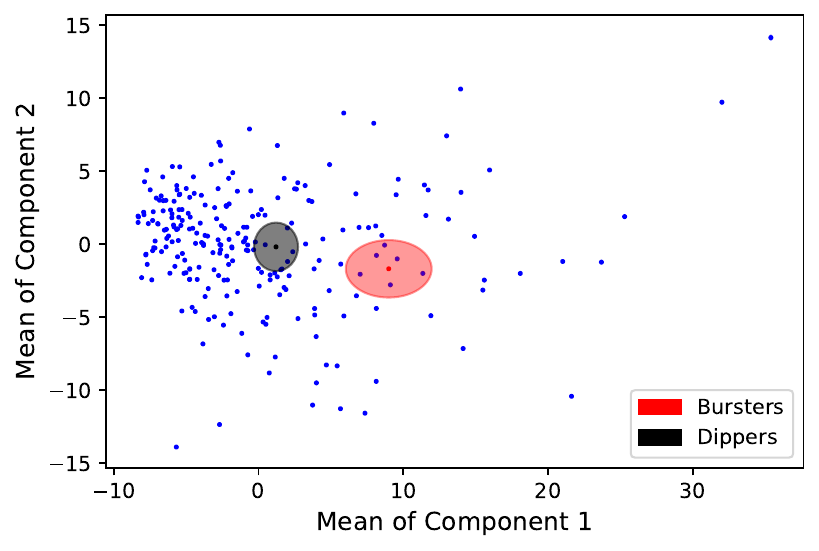}
\caption{\label{kmeans_bursterdipper} Placement test using PCA for our sample of 240 variable objects plus model light curves representing bursters and dipper. See text for details on the model parameters. The mean positions (points) and their standard deviations (shaded ellipses) for the models and data are determined in the same way as for Fig.~\ref{kmeans_sinewave}. }
\end{figure}

In addition to the simple sine waves, we briefly investigate quantitatively the placement of burster or dipper like light curves in the landscape. A more detailed, qualitative investigation of this will be included in the above mentioned follow-up investigation. We performed the same simulations (in terms of cadence, phase shifts, and repeats) as above, but changing the shape of the light curve into periodic dippers or bursters. The dippers are represented by symmetric dimming events, 0.4~mag deep. They are placed into the light curves every 150~d, with a duration of 2 months. The middle 60 percent of the duration are spent in the dim state, with the remainder in ingress and egress. Bursts are placed into the light curve as one magnitude brightness increases every two years with a duration of six months. The rise is short within 10~d, then there is a 2.5 month plateau before the brightness decreases towards the baseline.

In Fig.~\ref{kmeans_bursterdipper} we show the placement of these models in the landscape. We can see that their placement in the landscape and the scatter follow the same principles as for the sine waves. Models with more frequent events (in this case the dippers) are placed at lower PC~1 values and vice versa. The shape of the variation seems to have more of a subtle influence on the placement along PC~2. We note that this is just a proof of concept example. The exact locations and scatter of the model light curves will systematically depend on their parameters. Furthermore, these simple model dipper/burster light curves do not represent the full complexity of real YSO light curves. It should also be stressed that real objects coinciding with the locations of these models have the same variability statistics (fingerprints) but not necessarily light curves with the same dimming/brightening features as in these simplified models. 

\subsection{Placement in PCA generated landscape}\label{placement}

\begin{figure*}
\centering
\includegraphics[angle=0,width=\columnwidth]{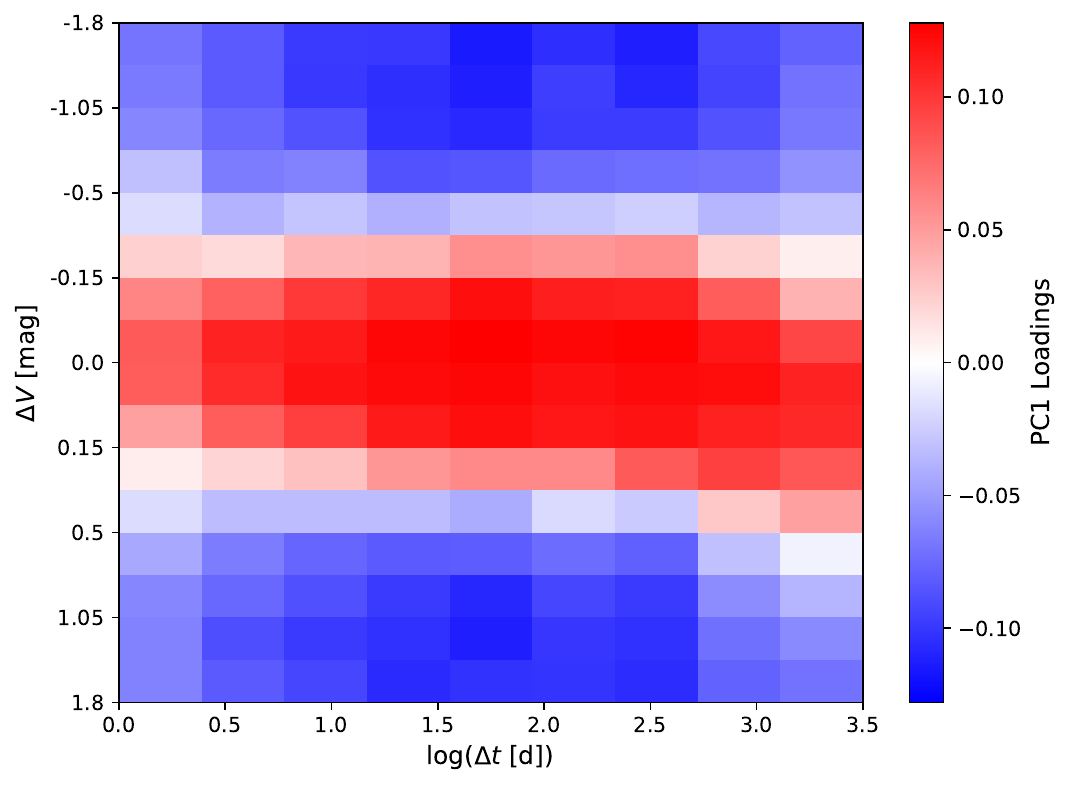} \hfill
\includegraphics[angle=0,width=\columnwidth]{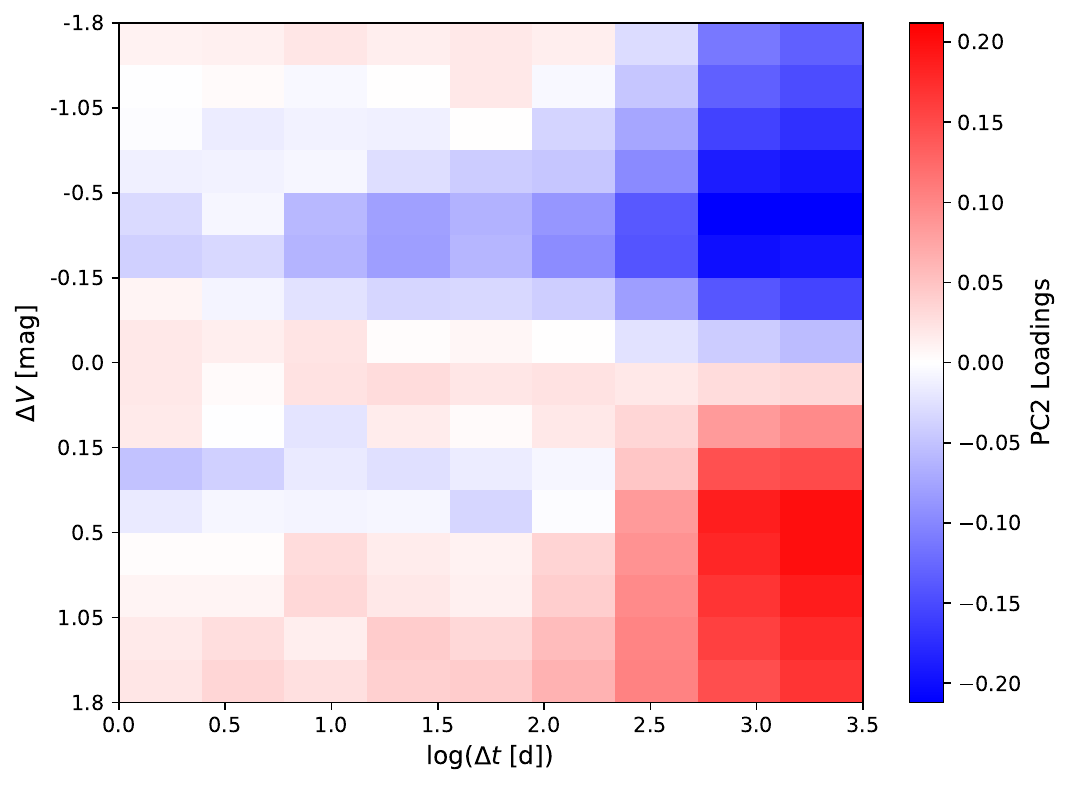}
\caption{\label{loadings_matrix} Loadings matrices of the 9~x~16 adaptive pixel fingerprints for principle component~1 (left) and principle component~2 (right). The colour scale shows how much each pixel value contributes to the positioning in the fingerprint landscape. }
\end{figure*}

As discussed in detail above, the distribution of objects within the fingerprint landscape forms a continuum rather than distinct clusters. Nevertheless, certain regions of the landscape are more densely populated, while some objects appear as clear outliers. This suggests that specific light curve characteristics are more prevalent than others. Furthermore, the artificial sine wave like light curves show a systematic trend in their placement which is governed by the period. Thus, we investigate here which light curve properties determine the placement of objects in the fingerprint landscape.

This is best done by visualising the loadings matrices for the two components generated by the PCA. These are shown in Fig.~\ref{loadings_matrix}. The colour code shows the weighting for each pixel in the fingerprints to the placement along each of the two components. In other words, if fingerprints have large pixel values in areas of the loading matrices that are red coloured, they will be placed towards larger values along the respective component (and vice versa). 

For PC~1 (left panel in Fig.~\ref{loadings_matrix}) we find that all fingerprint pixels that characterise light curve variations with an amplitude of less than 0.3~mag will push an object towards large values of PC~1. On the other hand, all variability in the light curve above an amplitude of 0.3~mag pushes the object towards smaller PC~1 values. Thus, a theoretically non-variable light curve would appear at very large PC~1 values. A high amplitude light curve with stochastic variations on very short time scales will be placed at very low PC~1 values. Light curves that start to vary on a given time scale will be placed in between the two extremes. 

We find that PC~1, derived from our sample of highly variable YSO light curves therefore primarily reflects the timescale for the onset of significant variability (exceeding 0.3~mag). This indicates that the timing of the onset of such variability serves as the most effective distinguishing feature among these light curves. The largest contributions to PC1, i.e. the highest absolute values in the loadings matrix, occur at timescales between one and three months. This range is consistent with the findings of \citet{2017MNRAS.465.3889R}, who reported that most Class~II YSOs exhibit low-amplitude (below 0.3~mag) optical variability driven by rotational modulation on timescales of a few weeks. In contrast, stronger, longer-term variability associated with the inner disk and accretion processes affects roughly one fifth of the population. Our results thus confirm that the onset of this additional, larger-amplitude variability typically occurs on timescales of one to three months.

The loadings matrix for PC~2 (right panel in Fig.~\ref{loadings_matrix}) is different. It is composed of pixel values close to zero for all time scales below about 1.5~yr. Thus, for the placement of an object along PC~2, only the long-term behaviour of the light curve matters. In particular, light curves that show long-term brightening (i.e. negative $\Delta V$ values) will be placed at lower PC~2 values. Light curves that show long-term dimming behaviour have higher values of PC~2.

The loadings matrices derived from the sample of YSO light curves remain qualitatively and quantitatively consistent across varying conditions. Specifically, the choice of filter, the resolution of the fingerprints, and the threshold applied to the variability indices have no impact on the interpretation of the loadings matrices. As a result, replacing a single object in the sample produces only negligible changes to the loadings matrices, accounting for the landscape stability described in previous sections. However, introducing a sample with substantially different characteristics, such as non-variable sources or a dominance of pure sine wave light curves, might alter the loadings matrices, potentially affecting the interpretation of PC~1/2.

The loading matrices for PC~1/2 can now also help us understand the systematic placement of sine wave like light curves in the landscape - as seen in Fig.~\ref{kmeans_sinewave} and discussed in Sect.~\ref{pca_artificial}. All these individual model light curves have been placed into the sample one by one. Thus, the loading matrices for each case are almost identical to those shown in Fig.~\ref{loadings_matrix}. The placement of the sine waves along PC~1 is hence easy to understand. The variability in these light curves starts at time scales of the order of a quarter of the period. Thus, the longer the period, the more non-variable does the light curve appear at shorter time scales, and the further it is moved towards higher PC~1 values. If we create the same plot as in Fig.~\ref{kmeans_sinewave} but use 0.5~mag amplitude sine waves, then the models are systematically placed towards larger PC~1 values compared to the 1~mag amplitudes. This again is understandable because more of the detectable variations will be at amplitudes below 0.3~mag, which cause the objects to move towards larger PC~1 values. Along PC~2 the sine wave models systematically move towards larger values with increasing period. The long-term behaviour of the sine waves should be symmetric. However, there is a subtle feature in the loadings matrix for PC~2. For all timescales below about 1.5~yr, variations of low amplitude cause a shift towards positive PC~2 values. The longer the period in the sine wave, the more likely we find low amplitude variations at shorter timescales, explaining the systematic shift. This feature is more evident in the higher resolution maps.

\section{Conclusion}

We have refined the treatment of photometry in our inhomogeneous HOYS dataset. Building on the standard calibration established in \citet{2018MNRAS.478.5091F} and \citet{2020MNRAS.493..184E}, we exclude unreliable photometry near bright stars and from images with tracking issues. Additionally, potential photometric outliers in colour-magnitude space are identified and removed from the analysis. Using our previous astrometric selection of $\approx$3000 members in the monitored young clusters \citep{2024MNRAS.529.1283F} and long-term $V$, $R$, and $I$-band photometry, we identify a sample of 240 highly variable YSOs.

We construct variability fingerprints from the light curves of these sources, mapping the probability that an object varies by a given amount over a given timescale. This enables a quantitative comparison of the variability statistics of stochastically varying sources with randomly sampled light curves. Our data allow us to probe variability from $\pm$~0.05~mag to $\pm$~2.0~mag on timescales from 1~d to 8.6~yr. Over 90 percent of the fingerprints achieve a signal-to-noise ratio above three, with low-S/N regions confined to short timescales and large amplitudes. Fingerprint uncertainties closely follow Poisson counting statistics.

Applying various dimension reduction and clustering techniques, we assess whether fingerprints form distinct groups and whether individual objects maintain stable relative positions when modifying the sample. Our analyses show that fingerprints create a continuous distribution rather than discrete clusters, consistent with a sample exhibiting variability from multiple physical processes, including accretion rate changes, line-of-sight extinction variations, and spotted stellar surfaces. However, objects close together in the landscape have similar light curve morphologies.

We find that dimension reduction via t-SNE does not yield a stable landscape due to its non-linear nature. Individual object positions shift significantly when modifying the sample, though the large-scale structure remains recognizable. Conversely, PCA produces a highly stable landscape where adding or altering a single object results in only marginal shifts of the remaining points. This stability allows us to incorporate model-generated fingerprints and assess their placement relative to observations.

To test this, we simulate simple sinusoidal and burster/dipper light curves with varying HOYS cadences and randomized phase shifts. These models occupy a limited region of the observed landscape, indicating that observing time, cadence, and photometric errors do not significantly affect an object's placement.

Analysis of the PCA loadings matrices shows that the primary source of variance among fingerprints derived from highly variable YSO light curves is the timescale at which significant variability (exceeding 0.3~mag) begins, with timescales of 1–3 months being the most prominent. The second most important contributor to the variance is long-term behaviour, specifically gradual fading or brightening trends occurring over timescales longer than 1.5 years.

These results demonstrate that PCA of variability fingerprints provides a robust, quantitative method to compare the variability statistics of observed YSO light curves with artificial light curves derived from models of the underlying physical processes.

\section*{Acknowledgements}

We would like to thank all contributors of observational data for their efforts towards the success of the Hunting Outbursting Young Stars project. HS is supported by the Science and Technology Facilities Council under grant number STFC Kent ST/Y509267/1.


\section*{Data Availability Statement}

The data underlying this article are available in the HOYS database at https://astro.kent.ac.uk/HOYS-CAPS/.


\bibliographystyle{mnras}
\bibliography{bibliography} 



\appendix

\begin{figure*}
\section{Example light curve pairs }
\centering
\includegraphics[angle=0,width=\columnwidth]{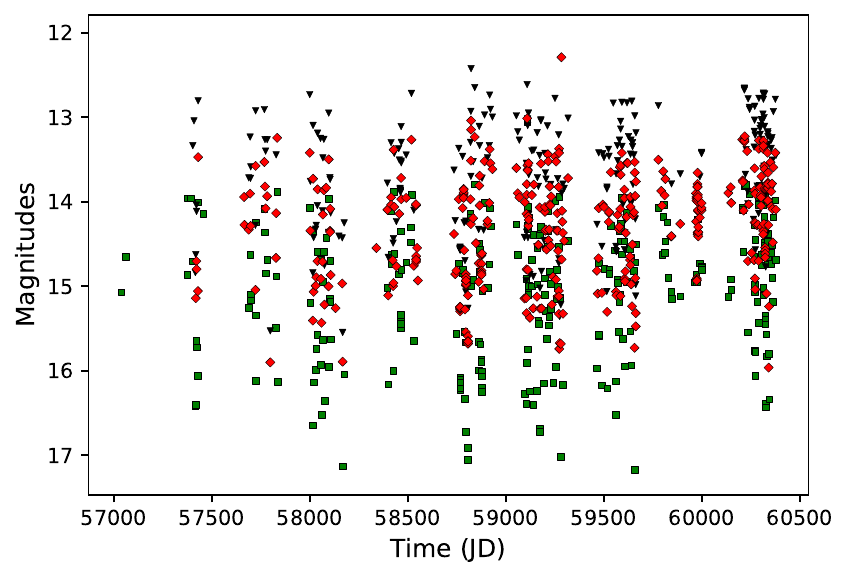} \hfill
\includegraphics[angle=0,width=\columnwidth]{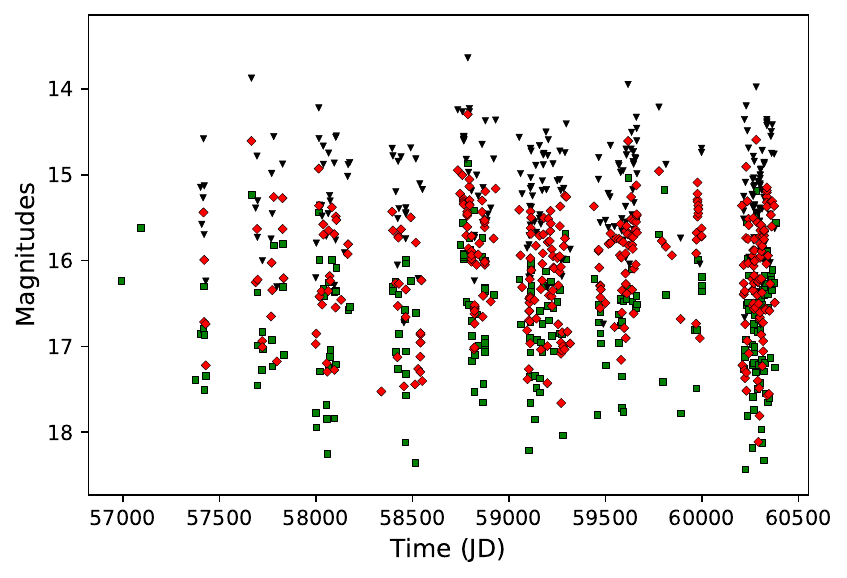} \\
\includegraphics[angle=0,width=\columnwidth]{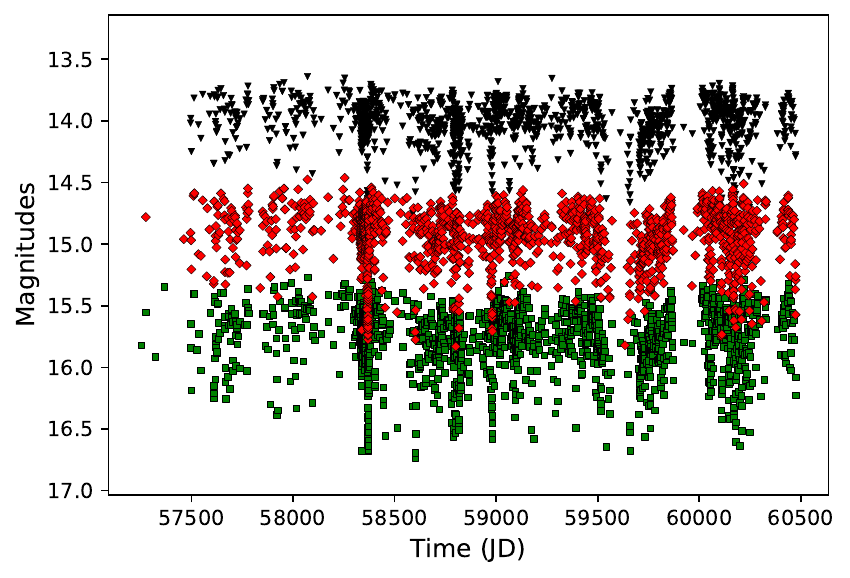} \hfill
\includegraphics[angle=0,width=\columnwidth]{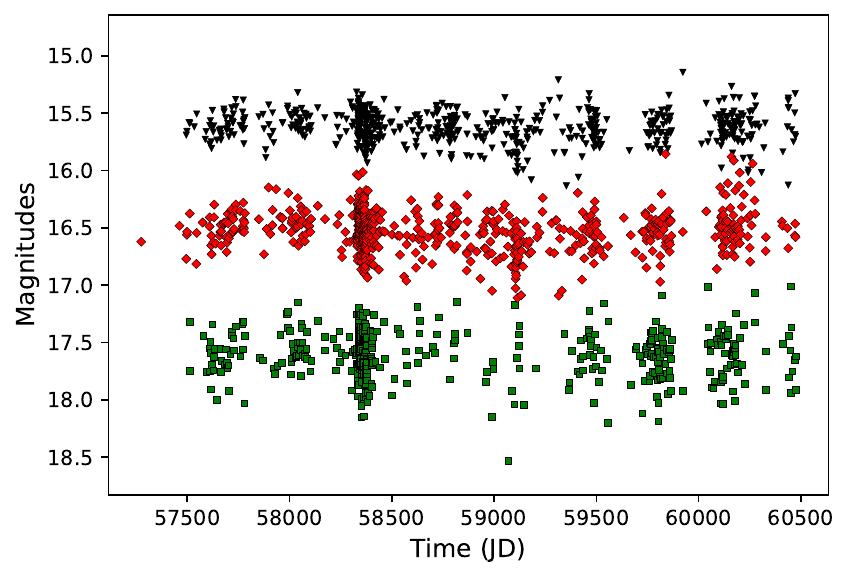} 
\includegraphics[angle=0,width=\columnwidth]{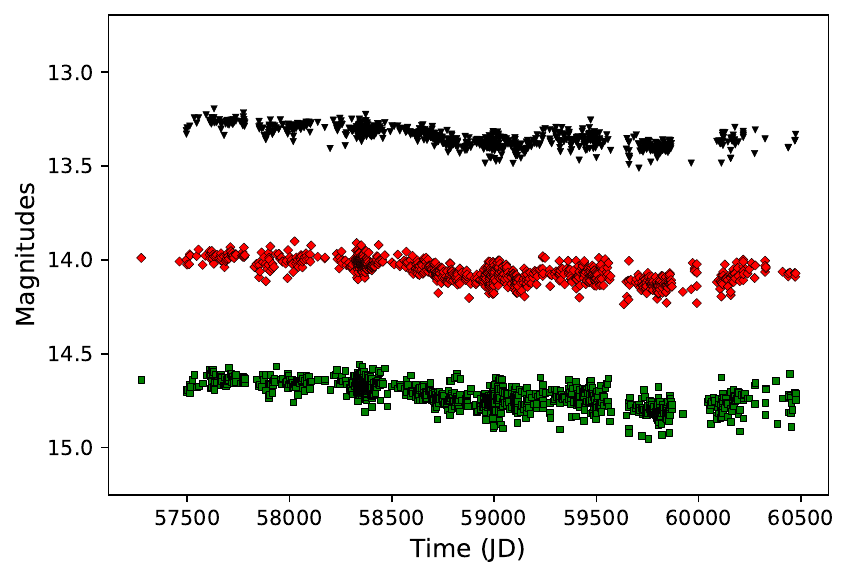} \hfill
\includegraphics[angle=0,width=\columnwidth]{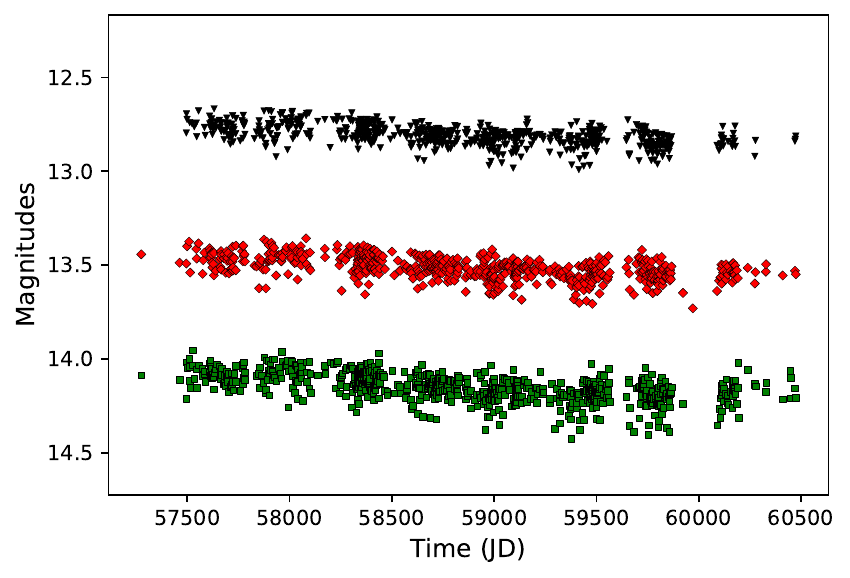} 
\caption{Example HOYS light curves. Each row contains two examples of light curves which are in close proximity to each other in the fingerprint landscape. In the {\bf top row} we have two sources with low PC1 values, the {\bf middle row} contains objects with intermediate PC1 values, and the {\bf bottom row} shows light curves with high values for PC1. \label{example_lcs}}
\end{figure*}

\bsp	
\label{lastpage}
\end{document}